\documentclass[10pt]{article}
\usepackage{graphicx}  
\usepackage{multirow}
\usepackage{array}
\usepackage{caption}

\usepackage{lipsum,multicol}

\usepackage{tabularx,ragged2e,booktabs,caption}

\usepackage{tikz}
\usepackage{pgfplots}
\usepackage{smartdiagram}

\usepackage{pdfpages}

\def\ind{{\mathchoice {\rm 1\mskip-4mu l} {\rm 1\mskip-4mu l}
{\rm 1\mskip-4.5mu l} {\rm 1\mskip-5mu l}}}
\usepackage{xcolor}

\begin{document}

\title{  Empathy in One-Shot Prisoner Dilemma}


\author{
        Giulia Rossi, Alain Tcheukam and Hamidou Tembine \thanks{  The authors are with Learning \& Game Theory Laboratory, New York University Abu Dhabi,
              Email:\ tembine@nyu.edu   }
}


\maketitle

\begin{abstract} 
Strategic decision making involves affective and cognitive functions like  reasoning, cognitive and emotional empathy which may be subject to age and gender differences. 
However, empathy-related changes in strategic decision-making and their relation to age, gender and neuropsychological functions have not been studied widely.  In this article, we study a one-shot prisoner dilemma from a  psychological game theory viewpoint.  Forty seven participants (28 women and 19 men), aged 18 to 42 years, were tested with a empathy questionnaire and  a one-shot prisoner dilemma questionnaire  comprising a closiness option with the other participant. The percentage of cooperation and defection decisions was analyzed. A new empathetic payoff model  was calculated to fit the observations from the test  whether multi-dimensional empathy levels matter in the outcome. A significant level of cooperation is observed in the experimental one-shot game.
The collected data suggests that perspective taking, empathic concern and fantasy scale are strongly correlated and  have an important effect on cooperative decisions. However, their  effect in the payoff is not additive. Mixed scales  as well as other non-classified subscales (25+8 out of 47) were observed from the data. 
 
\end{abstract}

\tableofcontents
\newpage 
\section{Introduction}

{\color{black}
 In recent years  a growing field of behavioral game studies has started to emerge from several academic perspective. Some of these  approaches and disciplines such  as neuroscience, social psychology, artificial intelligence have already produced major collections and experiments on empathy. In the context of strategic interaction, empathy may play a key role in the decision-making and the outcome of the game.
 
 \subsection*{ Widely known results in repeated games} 
 
 For long-run interactions under suitable 
  monitoring assumption it has been shown that  cooperative outcomes may emerge as time goes. This is known as  ``Folk Theorem" or general feasibility theorem (see  \cite{folk1,folk2}).   For example, the Tit-For-Tat Strategy which consists to start the game  by cooperating,Then do whatever your other participant  did on the previous iteration, leads to a partial cooperation between the players . While  cooperation may emerge by means of repeated long-run interactions under observable plays, there is very little study on how cooperation can be possible in one-shot games.
  
  \subsection*{ How about cooperative behavior in one-shot games?}
  
Unfortunately, the Folk theorem result does not apply to one-shot games.  This is because there is no previous iteration. There is no next iteration in one-shot games.  There is no opportunity to detect, learn or punish from experiences. For the same reasons, the existing reputation-based schemes do not apply directly. 

  \subsection*{Is cooperation possible outcome in experimental one-shot games?}
Experimental results have revealed a strong mismatch between the outcome of the experiments and the predicted outcome from classical game model. Is the  observed mismatch because of  some important factors that are not considered in the classical game formulation? Is it because the empathy of the players are neglected in the classical formulation?

This work conducts a basic experiment on one-shot prisoner dilemma and establishes correlation between players choices and their empathy levels. 
The prisoner's dilemma is a canonical example of a game analyzed in classical game theory that shows why two individuals (without empathy consideration) might not cooperate, even if it appears that it is in their best interests  to do so in terms of collective decision. It was originally framed by Flood and Dresher working at RAND. In 1950, Tucker gave the name and interpretation of prisoner's dilemma to Flood and  Dresher's model of cooperation and conflict, resulting in the most well-known game theoretic academic example.}

\subsection*{Contribution}

The contribution of this work can be summarized as follows. 
We investigate how  players behave and react in experimental one-shot Prisoner Dilemma in relation to their levels of empathy. The experiment is conducted with several voluntary participants from different countries, cultures and educational backgrounds. For each participant to the project the Interpersonal Reactivity Index (IRI)  which is a multi-dimensional empathy measure, is used. In contrast to the classical empathy scale studied  in game theory literature that are limited to perspective taking, this work goes one-step further by investigating the effect of three other empathy subscales: empathy concern, fantasy scale and personal distress.  The experiment reveals a strong mixture of the empathy scales across the population.   In addition, each participant responds to a questionnaire that mimics  one-shot Prisoner Dilemma situation and specific reaction time and closiness to the other participant. 
We observe that  empathic concern as well as fantasy scale dimensions  may affect positively other-regarding payoff. 
In contrast to the classical prisoner dilemma in which Defection is known as a dominating strategy, the experimental game exhibits a significant  level of cooperation in the population (see Section \ref{newmodel1}). 
In particular, Defection
 not a dominating strategy anymore when players' psychology is involved.   
  Based on these observations an empathetic payoff model in Section \ref{newmodel}, that better captures the preferences of the decision-makers, is proposed. With this empathetic payoff, the outcome of the game  captures more the observed cooperation level in the one-shot prisoner dilemma.
The experiment  reveals  not only positive affect of empathy but also a dispositional negative affect  (spiteful or malicious) of empathy in the decision-making of some of the participants. Spitefulness is observed at the personal distress scale in the population. Person distress scale is negatively correlated with perspective taking scale.
  Taken together, these findings suggest that the empathy types of the participants play a key role  in their payoffs and in their decision in one-shot prisoner dilemma.  It also  reduces the gap  between game theory and game practice by closing-the-loop between model and observations. It provides an experimental evidence that  strengthens the model of empathetic payoff  and its  possible  engineering applications \cite{t1}.

\subsection*{Stucture}
The rest of the article is organized as follows.  The next section presents some background and literature overview  on empathy. Section 3  presents the experimental study about the impact of individual psychology on human decision making in one-shot prisoner dilemma.  The analysis of the results of the experiment  are presented in Section 4.  An explanation of the results  is given in Section 5.  Section 6 concludes the paper.


\section{Background on Empathy}

From the field of relational care to the field of economy,  the concept of empathy seems to have an ubiquitously position. Empathy is not only an important and longstanding issue, but a commonly used term in everyday life and situations.  Even if it is easily approached and used, this concept has until nowadays different definitions and meanings. Born in the aesthetic and philosophical field, it came to be an important operative concept in behavioral game theory, where once more it is used 
as an instrument to create relations between decision-makers.    The public opinion and the world scientific scenario, however, are not  always giving the correct attention to what is implying an empathetic reaction with the others: be empathetic is not something simple as well as is not something given once to the personality of persons. It depends on the context of relations and on the social interaction dimension where people are involved as players, consumers, or agents.

\subsection*{Definition of Empathy }
{\color{black} We present  historical definitions and concepts of empathy. Rather than having to choose which of the 'definitions' of empathy is correct, this work suggests a better appreciation for it as a multidimensional phenomenon at least allows a perspective and the ability to specify which aspect of empathy the experimentalist and the theorist are referring to when making particular particular investigation in behavioral games.}

\begin{itemize}\item 
{\color{black}Philosophy:} From philosophical perspective, Empathy derives from the Greek word $\acute{ \epsilon } \mu  \pi  \acute{\alpha}   \theta   \epsilon i    \alpha$ (empatheia), which literally means physical affection. In particular, it is composed by $\acute{ \epsilon} v$ (en) ``in, at" and $ \pi   \acute{\alpha} \theta  o \varsigma $ (pathos), ``passion'' or  ``suffering''. The work in  \cite{c1} has introduced the term Einfuhlung to aesthetic philosophical field in his main book 
``On the optical Sense of Form: A contribution to Aesthetics'' and many other authors (see   \cite{c2} and the references therein), have introduced the concept to feeling and quasi-perceptual acts. 
\item 
{\color{black}Psychology:} From  a psychological perspective, it corresponds to a cognitive awareness of the emotions, feelings and thoughts of the other persons. In this sense, the term primary significance is that of  ``an intellectual grasping of the affects of another, with the result of a mimic expression of the same feelings''  \cite{c3}. 
\item 
{\color{black}Sociology: }From a sociological perspective, empathy corresponds to an ability to be aware of the internal lives of the others. It is related to the existence of language as a sort of 
personal awareness of us as selves  \cite{c4}. Regarding a neuroscientific perspective, empathy has been studied as a 
 ``two empathic sub-processes, explicitly considering those states and sharing other's internal states''  \cite{c5}. These two cognitive processes are exhaustively represented in the Figure \ref{fig1}.

%

\begin{figure}[htbp]
\begin{center}
\centering
\includegraphics[width=8cm,height=4cm]{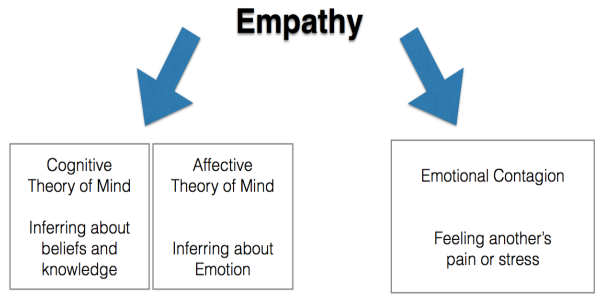}
\caption{Cognitive and emotional empathy are the bases of Empathy. They are related to cognitive and affective Theory of Mind as suggested in   \cite{c10}.}
\label{fig1}
\end{center}
\end{figure}

Empathy lies in the structures that include the anterior insula cortex and the dorsal anterior cingulate cortex. In particular, empathy for others pain is considered being located in anterior insula cortex \cite{c7,c6}. These areas are studied in relation to empathy and empathy concerns of collective actions (CA)  \cite{c8}. Oxytocin (OT), an ``evolutionarily ancient molecule that is a key part of the mammalian attachment system'', is considered in these studies as a sort of variable to be manipulated to increase or decrease CA in human beings.


\end{itemize}	 
	 
Furthermore, empathy must be analyzed in relation to the concepts of compassion and sympathy. Compassion, from Latin ecclesiastical compati (suffer with), literally means have feelings together. Nowadays it is associated with the capacity of feeling the other's worries, even tough it doesn't imply an automatic action. Sympathy, from the Greek  $ \sigma  u  \mu  \pi  \acute{a}   \theta   \epsilon i a     $, literally means fellow feelings. Its meaning lies in the capacity of understanding the internal feelings of the other with the intentional desire of changing his/her worries. As per indication in \cite{c9},  ``the object of sympathy is the other person's well being. The object of empathy is understanding''.  It has been difficult to distinguish empathy from sympathy because they both involve the emotional state of one person to the state of another. This problem was increased by the fact that the mapping of the terms has recently reversed: what is now commonly called empathy was referred to before the middle of the twentieth century as sympathy \cite{c10ptreston}.
At the end, the concept of empathy must be understood also in relation to Theory of Mind. Theory of Mind, also described as ToM, is the capability to understand of others as mental beings, with personal mental states, for example feelings, motives and thoughts.  It is one of the most important developments in early childhood social cognition and it is influencing children life at home as well as at school. Its development from birth to 5 years of age is now described in research literature with the possibility to understand how infants and children behave in experimental and natural situations \cite{d4}.

\subsection*{Different types of empathy }
We are not restricting ourselves to the positive part of empathy. Empathy may have a dark or at least costly side specially when the environment is strategic and interactive as it is the case in games. Can empathy be bad for the self? Empathy can be used, for example, by a other player attacker to identify the weak nodes in the network.
Can empathy be bad for others? Empathetic users may use their ability to destroy the opponents. In strategic interaction between people, empathy may be used to derive antipathetic response (distress at seeing others' pleasure, or pleasure at seeing others distress). In both cases, it will influence the dynamics of the self and other regarding preferences.


The ability of empathy to generate moral behavior and determine cooperation is limited by three common occurrences: over-arousal, habituation and bias.                                                                                                                                                                                                                                                                                                                                                                                                                                                                                                                                      
\begin{itemize}\item Empathic over-arousal is  an involuntary process that occurs when an observer's empathic distress becomes so painful and intolerable that it is transformed into an in tense feeling of personal distress, which may move the person out of the empathic mode entirely \cite{f13,f14} . 
\item Generally speaking, in 
a classical relation victim-observer, the greater is the victim's distress, the greater is the observer's empathic distress. If a person is exposed repeatedly to distress over time, the person's empathic distress may diminish to the point where the person becomes indifferent to the distress of others. This is called habituation. This diminished empathic distress and corresponding indifference is very common in those who, for example, abuse and kill animals.
\item  Humans evolved in small groups. These groups sometimes competed for scarce resources: in this way is not surprising that evolutionary psychologists have identified kin selection has a moral motivator with evolutionary roots. The forms of familiarity bias include in-group bias and similarity biases. In-group bias is simply the tendency to favour one's own group. This is not one group in particular, but whatever group we are able to associate with at a particular time. In-group bias is working on self-esteem of the members.  On the opposite side of these biases, we have out-group ones, where people out of the groups are considered in a negative way, with a different (and, for the most of the time) and worst treatment (e. g, racial inequality). The similarity bias derives from psychological heuristic pertaining to how people make judgments based on similarity. More specifically, similarity biases are used to account for how people make judgments based on the similarity between current situations and other situations or prototypes of those situations. The goal of these biases is to maximize productivity through favorable experience while not repeating unfavourable experiences (adaptive behaviour).
\end{itemize}

\subsection*{Empathy main integrative theories }

During the 1970s, in the public psychological scenario empathy was conceived as a procedure with affective and cognitive implications.  The work in \cite{d12}  introduced the first multidimensional model of empathy in the psychological literature, where affective and cognitive procedures were working together. According to her, although empathy is defined as a shared emotion between two persons, it depends on cognitive factors. In her integrative-affective model, the affective empathy reaction derives from three components factors, hereinafter described. The first is represented by the cognitive ability to discriminate affective cues in others, the second by  the cognitive skills that are involved in assuming the perspective and role of the others and the third factor is, at the end, described by  the emotional responsiveness, the affective ability to experience emotions.
By the other hand, one of the most comprehensive perspectives on empathy and its relation to the moral development is provided in \cite{d13}. The author  considered empathy as a biologically based disposition for altruistic behavior \cite{d13}.  He conceives of empathy as being due to various modes of arousal, which allows us to respond empathically in light of a variety of distress cues from another person. The author mentions mimicry, classical conditioning, and direct association as fast acting and automatic mechanisms producing an empathic response. The author lists mediated association and role taking in relation to more cognitively demanding modes, mediates by language and proposed some of the limitations in our natural capacity to empathize or sympathize with others, particularly what he refers to as 'here and now' biases. In other words, the main tendency according to \cite{d13}, is to empathize more with persons that are in some sense perceived to be closer to us. 
The authors in \cite{prest02}    defined empathy as a shared emotional experience occurring when one person  (comes to feel a similar emotion to another (the object) as a result of perceiving the other's state. This process results from the representations of the emotions  activated when the subject pays attention to the emotional state of the object. The neural mechanism assumes that brain areas have processing domains based on their cellular composition and connectivity.

The main theories that we will discuss further are related to the use we will have of empathy concept in game theory analysis. Broadly speaking, we are approaching empathy as made up of two components, an affective and a cognitive one. The affective (or emotional) component develops from infants and its structure may be summarized in this progressive intertwinement, a.) emotion recognition, b.) empathic concern, c.) personal distress and d.) emotional contagion. The cognitive component of empathy develops during the progressive development of the person from his \ her childhood. It is based on the theory of mind, imagination (of emotional future outcomes) and on perspective taking.

According to  \cite{d5} two main approaches have been used to study empathy: the first one focuses on cognitive empathy or the ability to take the perspective of another person and to infer his mental state (Theory of Mind). The second one emphasizes emotional or affective empathy \cite{d6} defined as an observer's emotional response to another person's emotional state. The Table \ref{fig3} below highlights some of  principal features of affective and cognitive empathy \cite{c10}.
\begin{table*}
\begin{tabular}{ |c|c|  }
  \hline
Emotional Empathy& Cognitive Empathy \\ 
  \hline 
  Simulation System  & Mentalizing system  \\
  & Theory of Mind  \\ \hline
  Emotional contagion, personal distress & perspective taking\\
 Empathic concern, emotion recognition & imagination of emotional future outcomes\\
  Core structure & Core structure\\
  Development & Development\\
  \hline
\end{tabular} \caption{Principal features of affective and cognitive empathy.} \label{fig3} 
\end{table*}

A model of empathy-altruism was developed in \cite{b8}. Through the lines of this model, the author simply assumes that empathy feelings for another person create an altruistic motivation to increase that person's welfare. In particular, the work in \cite{c11ba} found out how the participants in a social dilemma experiment allocated some of their resource to a person for whom they felt empathy. The author developed also the model of empathy-joy \cite{b8}. This hypothesis underlines how a prosocial act is not completely explained only by empathy but, also, by the positive emotion of joy a helper expects as a result of helping  another person in need. In connection with this theory, empathy relies on an automatic process that generates, immediately, other types of behavior useful to predict the other-regarding actions. In relation to one-trial prisoner's dilemma \cite{c12} underlined how empathy-altruism should increase cooperation (that, then, will emerge in the situation).

The model of empathic brain proposed in \cite{c13} proposes a modulate model of empathy, where different factors occur in its development.  These factors are four, in particular related to four different situations so described, i) one is in affective state, ii) this state is isomorphic to another person's affective state, iii) this state is elicited by the observation or imagination of another person's affective state, iv) one knows that the other person is the source of one's own affective state. Condition (a) is particularly important as it helps to differentiate empathy from mentalizing. Mentalizing is the ability to represent others' mental states without emotional involvement. In particular, the two authors are underlying the epistemological value of empathy, by one side related to provide information about future actions between people and, by the other side, to function as ``origin of the motivation for cooperative and prosocial behavior''. The model we provide is based on both these theories and, in particular, it will take into account the distinction between empathy itself and the cortical representations of the emotions. By a developmental point of view, empathy is studied also in relation with prosocial behavior. Two theoretical studies  \cite{d1,d2}    are fundamental in this sense.  In particular, the author considers that the development of a vicarious affective reaction to another distress is beginning from infancy. Individual patterns of behavior that responses to the distress are, also, tailored to the needs of the other. According to \cite{d3}  although an empathic basis to altruistic behavior entails a net cost to the actor, cooperation and altruism require behavior tailored to the feelings and needs of others.

\subsection*{How to measure empathy?}
We overview empathy measurement in psychology and  present existing models of empathy's effect in game theory.

\subsubsection*{Empathy  measures in Psychology}

Psychologists used to study both situational and dispositional empathy concepts.  Situational empathy, i.e., empathic reactions in a specific situations, is measured by asking subjects about their experiences immediately after they were exposed to a particular situation, by studying the ``facial, gestural, and vocal indices of empathy-related responding"  \cite{d7} or by various physiological measures such as the measurement of heart rate or skin conductance. Dispositional empathy, understood as a person's stable character trait, has been measured either by relying on the reports of others (particularly in case of children) or, most often (in researching empathy in adults), by relying on the administration of various questionnaires associated with specific empathy scales.
\begin{itemize} \item {\color{black} Measuring empathic ability:}
The work in \cite{d8} proposes to test empathic ability by measuring the degree of correspondence between a person A and a person B's ratings of each other on six personality traits-such as self-confidence, superior-inferior, selfish-unselfish, friendly-unfriendly, leader-follower, and sense of humor-after a short time of interacting with each other. More specifically, empathic ability is measured through a questionnaire that asked both persons, i) to rate themselves on those personality traits, ii) to rate the other as they see them, iii) to estimate from their perspective of how the other would rate himself and to rate themselves according to how they think the other would rate them. Person's A empathic ability is then determined by the degree to which A's answers to (iii) and (iv) corresponds to B's answer to (i) and (ii). The less A's answers diverge from B's, the higher one judges A's empathic ability and accuracy. The test aims to measure the level of empathy thanks to the dimension of role-taking.
\item {\color{black} Empathy Test:}
The authors in  \cite{d9} created the so-called Empathy Test, that was used in industry in the 1950s.  The main purpose of the test is to measure person's ability to ''anticipate'' certain typical reactions, feelings and behaviour of other people. This test consists of three sections, which require persons to rank the popularity of 15 types of music, the national circulation of 15 magazines and the prevalence of 10 types of annoyance for a particular group of people.
\item {\color{black} Measure of  cognitive empathy:}
The authors in  \cite{d10} is a cognitive empathy scale that consists of 64 questions selected from a variety of psychological personality tests such as the Minnesota Multiphasic Personality Inventory (MMPI) and the California Personality Inventory (CPI). Hogan chose those questions in response to which he found two groups of people-who were independently identified as either low-empathy or high-empathy individuals-as showing significant differences in their answers.
\item {\color{black} Measure of emotional empathy:}  EETS,  Emotional Empathy Tendency Scale, has been developed in \cite{d11}.
The questionnaire consists of 33 items divided into seven subcategories testing for ``susceptibility to emotional contagion", ''appreciation of the feelings of unfamiliar and distant others,``extreme emotional responsiveness",  
``tendency to be moved by others' positive emotional experiences",   ``tendency to be moved by others' negative emotional experience",  ``sympathetic tendency", and ``willingness to be in contact with others who have problems". This questionnaire emphasizes the original definition of the empathy construct  in its seven subscales that together show high split-half reliability, indicating the presence of a single underlying factor thought to reflect affective or emotional empathy. The authors in \cite{f1} suggested more recently, however, that rather than measuring empathy per se, the scale more accurately reflects general emotional arousability. In response, a revised version of the measure, the Balanced Emotional Empathy Scale   \cite{f2} intercepts respondent's reactions to others' mental states \cite{f3}.
\item {\color{black} Multidimensional measure of empathy:} The Interpersonal Reactivity Index has been developed in \cite{d18} as an instrument whose aim was to measure individual differences in empathy.  The test is made of 28 items belonging to cognitive and emotional domain, and, in particular, they are belonging to four different domains. These are represented by four different subscales, Perspective Taking, Empathic Concern, Fantasy and Personal Distress, each of which includes seven item answered on a 5-point scale ranging from 0 (Does not describes me very well) to 4 (Describes me very well). The Perspective Taking subscale measures the capability to adopt the views of the others spontaneously. The Empathic Concern subscale measures a tendency to experience the feelings of others and to feel sympathy and compassion for the unfortunate people.
\item  {\color{black} Self-report empathy measure:}
Regarding the dimension of  self-report empathy measures, we consider important to be mentioned the Scale of Ethnocultural Empathy  \cite{f4}, the Jefferson Scale of Physician Empathy   \cite{f5}, the Nursing Empathy Scale \cite{f6}, the Autism Quotient  \cite{f7} and the Japanese Adolescent Empathy Scale  \cite{f8}. Although these instruments were designed for use with specific groups, aspects of these scales may be suitable for assessing a general capacity for empathic responding. 
\item  {\color{black} Measuring deficit in theory of mind:}
The Autism Quotient  \cite{f7} was developed to measure Autism spectrum disorder symptoms. The authors  viewed a deficit in theory of mind as the characteristic symptom of this disease  \cite{f9} and a number of items from this measure relate to broad deficits in social processing (e.g., ``I find it difficult to work out people's intentions.''). Thus, any measure of empathy should exhibit a negative correlation with this measure. The magnitude of this relation, however, will necessarily be attenuated by the other aspects of the Autism Quotient, which measure unrelated constructs (e.g., attentional focus and local processing biases).
Additional self-report measures of social interchange appearing in the neuropsychological literature contain items tapping empathic responding including the Disexecutive Questionnaire   \cite{f10} and a measure of emotion comprehension developed  in \cite{f11}. These scales focus on the respondent's ability to identify the emotional states expressed by another (e.g., ``I recognize when others are feeling sad".). Current theoretical notions of empathy emphasize the requirement for understanding of another's emotions to form an empathic response \cite{f12}. Only a small number of items on current measures of empathy, however, assess this ability. Table \ref{table:Empathymeasurement}  summarizes  the Empathy scales measurement in Psychology overviewed above.
\end{itemize}

\begin{table*}
\begin{center}
\begin{tabular}{ | p{3cm} |   p{1cm} |    p{2cm} |     p{3cm}|    p{3cm} | } 
 \hline

  &  ITEMS  &  SCALE &  CORE CONCEPT   &   {\color{black}  AUTHORS}  \\ \hline
  
 Empathy  Ability   &  24   &  0-4 Likert Scale  & Imaginative transposing of oneself  into the thinking of another & Dymond (1949)    \\ \hline
 
 Empathy Test   &	40   &	Ranking Multiple Choice   & 	Cognitive role-taking   & 	Kerr \& Speroff (1954)   \\ \hline
 
Empathy Scale  &  	64	 & 0-4 Likert Scale  &	Apprehension of another's condition or state of mind by an intellectual or imaginati ve point of view  & Hogan (1969)  \\ \hline
 
 EETS - Emotional Empathy Tendency Scale   & 	33   & 	-4 to 4 Likert Scale  & 	Emotional Empathy	   &  Mehrabian \& Epstein (1972)   \\ \hline
 
 IRI- {\color{black} Interpersonal Reactivity Index}   &  	28	 &   0-4  Likert Scale	   &   Reactions of one individual to the observed experiences of another	   &  Davis (1980)   \\ \hline

 Ethnocultural Empathy Scale  & 	31 &  	& 	Culturally Specific Empathy Patterns	  &  Wang, David son,Yakushko, Savoy, Tan, Bleier (2003)   \\ \hline
 
 Jefferson Scale of Physician Empathy    &  	5    &  	0 - 5   Likert Scale   &  	Patient oriented vs technology oriented empathy in physicians   &  	Kane, Gotto, Mangione, West , Hojat, (2001)   \\ \hline
   
 Nursing Empathy Scale	 &    12	 &    0-7 Likert Scale	 &    Nurses empathic behaviour in the context of interaction with the client  &    Reynolds (2000)   \\ \hline
  
 Autism Quotient	  &  50   &    0-4  Likert Scale   &  	 Autism Spectrum Disorder symptoms	   &   Baron-Cohen, Wheelwright, Skinner, Martin, Cubley (2001)  \\ \hline

  Japanese Adolescent  Empathy Scale   &	30  &  	Likert Scale   &   	Empathy to feel or not to feel positive and negative feelings towards the others	&   Hashimoto, Shiomi (2002)  \\  \hline 
\end{tabular}
\end{center}
\caption{Empathy measurement in psychology}
\label{table:Empathymeasurement}
\end{table*}

\subsubsection*{Empathy Models in Game Theory}
In this subsection, we review some fundamental aspect of  behavioral game theory. By a game theoretic approach, empathy and emotive intelligence are considered essential for the development of the games themselves between players.  In particular, empathy is essential for the strategic evolution of the games and foundational of Nash Equilibrium  \cite{b2,b3}. In other words, empathy itself is the instrument that let the dynamic process between n-players happen as well as the understanding and evaluation of their preferences and beliefs.   Cooperative behavior patterns must be considered an important sample of close relations between individuals, based on confidding and disclosure. Relations like helping and assistance behavior, as well as mutual confiding, mutual communication and self disclosure are of cooperative behavior. To understand the possible role of friendship in cooperation or defection between n-players in one-shot prisoner dilemma, it's essential to reflect on how the evolutionary line of our specie has the possibility to create close relationship only between  persons who are considering themselves keen in terms of genes. The same kind of attitude is also influencing communal relationship.The cooperative behaviour in one shot Prisoner Dilemma between friends, in particular, lead to the activation of the so called cooperative-parithetic system, that is activated only when there is the perception that the last goal may be reached through a sort of collaboration between the players of the group itself. Empathy has been approached in different ways regarding game theory field.   The author in \cite{b4} underlines how homo economicus must be empathetic to some degree, even if in a different meaning from the concept used in \cite{d19,d20}. In particular, in relation to game theory, he introduces the concept of empathy in connection to the study of interpersonal comparison of utility in games. 

More specifically, the model of empathy - altruism developed in \cite{b8} whose assumption is that empathy feelings for another person create an altruistic motivation to increase that person's welfare.  Furthermore,  the work of \cite{d21}  is related to how the participants in a social dilemma experiment allocate some of their resource to a person for whom they felt empathy. In the context of one  trial prisoner's dilemma the authors  underlined how empathy altruism should increase cooperation (that, then, will emerge in the situation).
Secondly, the model of empathy  joy developed in \cite{d22}. This hypothesis underlines how a prosocial act is not completely explained only by empathy but, also, by the positive emotion of joy a helper expects as a result of helping (or, better, of having a beneficial impact on) another person in need.  In connection with this theory, empathy relies on an automatic process that generates, immediately, other types of behaviour useful to predict the other regarding actions \cite{book,rmg}.

The works in \cite{d15,d16} proposed the exploration of more psychological and process-oriented models as a more productive framework in comparison with the classical ones in game theory (fairness, impure altruism, reciprocity). At the light of this perspective, many concepts related to human behaviour are introduced to explain choice in game theory approach. Empathy as an operative concept must be understood as different from biases affecting belief formation and biases affecting utility. Empathy is operating through both beliefs and utility formation. Hence, in the presence of empathy,  ``beliefs and utility become intricately linked''  \cite{d17}. Regarding empathy as a process of beliefs formation, the author proposed to analyze two mechanisms, imagine-self and imagine-other.  Imagine self-players are able to imagine themselves in other people's shoes, in other words they try to imagine themselves in similar circumstances. ''Imagine other'' is when a person tries to imagine how another person is feeling. The authors underlined how empathy refers to people's capability to infer what others think or feel, the so-called mind reading. They underlined, at the same time, how empathy itself may have also any consequences on each player evaluation. The main contribution of the authors lies in a critique analysis of altruistic behaviour in game theory. With three toy games, they demonstrated how empathy-altruism is not always linked with imagine-other dimension (and the so-called beliefs formation), since players may use only imagine-self dimension.

The authors in \cite{d14} criticize a common concept of given empathy present in public good experiment and, at the end, they demonstrate how empathy may be linked more to the context and social interaction itself in game theory experimental researches.  The work \cite{b6} states that a disposition for empathy does not influence the behaviour related to different games (towards them a central role is played by Theory of Mind). Regarding this position, the same authors are underlying that also individual differences related to empathy do not shape social preferences. On the contrary, many other studies conducted show how empathy may influence the structure of the games themselves. The work in \cite{b7}  study the Ultimatum Game in an evolutionary concept and underline, in their study, how empathy can lead to the evolution of fairness. The work in \cite{b9} studied the correlation between empathy, anticipated guilt and pro social behaviour; in this study he found out that empathy affects pro-social behaviour in a more complex way than the one represented by the model of social choices.

Recently, the concept of empathy has been introduced  in mean-field-type games in \cite{t1,tt1v,tt2v,tt3v} in relation to cognitively plausible explanation models of choices in wireless medium access channel and mobile devices strategic interaction. The main results of this applied research that lie in an operative and real world use of empathy concept, are represented by the enforcement of mean-field equilibrium payoff equity and fairness itself between players.

\section{Study}
\subsection*{Participants}
The population of   participants includes 47 persons  between  18 and 42 years old.  The population is composed of 19 men and 28 women chosen from different educational backgrounds, cultures and nationalities (see Table \ref{table:gender}). The names of the participants are not revealed. Different numbers are generated and assigned to the participants.

\begin{table}[htb]
\centering
\begin{tabular}{|l|lr|} \hline
\cline{2-3}
Gender & Number &  Frequency \% \\ \hline
Men   & 19 & 40.42  \\ \hline
Women     & 28    &   59.58  \\ \hline
   Total       & 47        &       \\ \hline
\end{tabular}
\caption[]{ Composition:  gender and frequency of the participants }
\label{table:gender}
\end{table}

All the subjects were asked to perform two different tests: an  IRI test (Interpersonal Reactivity Index \cite{b11}) and   a questionnaire that is mimicking, with an empathic and moral emphasis, a prisoner dilemma situation. 

\subsection*{Empathy questionnaire} The IRI is a 28-item, 5-point Likert-type scale that evaluates four dimensions of empathy: Perspective-Taking, Fantasy, Empathic Concern, and Personal Distress.
Each of these four subscales counts 7 items. The Perspective-Taking subscale measures empathy in the form of individual's tendency to adopt, in a spontaneous way, the other's points of view. The Fantasy subscale of the IRI evaluates the subject's ability to put themselves into the feelings and behaviours of fictional characters in books, movies, or plays. The Empathic Concern subscale assesses individual's feelings of concern, warmth, and sympathy toward others. The Personal Distress subscale measures self-oriented anxiety and distress feelings regarding the distress experienced by others. As pointed out by Baron-Cohen and colleagues  \cite{d24}, however, the Fantasy and Personal Distress subscales of this measure contain items that may more properly assess imagination (e.g., ``I daydream and fantasize with some regularity about things that might happen to me'') and emotional self-control (e.g., ``in emergency situations I feel apprehensive and ill at ease"), respectively, than theoretically-derived notions of empathy. Indeed, the Personal Distress subscale appears to assess feelings of anxiety, discomfort, and a loss of control in negative environments. Factor analytic and validity studies suggest that the Personal Distress subscale may not assess a central component of empathy \cite{d25}. Instead, Personal Distress may be more related to the personality trait of neuroticism, while the most robust components of empathy appear to be represented in the Empathic Concern and Perspective Taking subscales \cite{d26}
 
IRI Davis Scale has been chosen for its relation to the measurement of individual differences in empathy construct and, secondly, for its relation with measures of social functioning and the so-called psychological superior functions  \cite{d23}.  Table \ref{tab:table1} summarizes the first questionnaire on multidimensional empathy measure.

\begin{table*}
\centering
 \caption{IRI subscales. 
Extension of the empathy measure of Davis 1980,   Yarnold et al.1996  and Vitaglione et al. 2003. The star sign (*) denotes an opposite (reversed) counting/scoring.}
  \label{tab:table1}
  \begin{tabular}{l|cccc|cccc|}
    \hline
    \multirow{1}{*}{Abridged item} &
      \multicolumn{4}{c}{Women (59.58\%)} &
      \multicolumn{4}{c}{Men(40.42\%)} \\
      & {PT} & EC& {FS} & {PD} &  {PT} & EC& {FS} & {PD}\\
      \hline
     (1) Daydream and fantasize (FS) &  &  &  &  & & & &  \\
    (2) Concerned with unfortunates (EC) &  & 0.6 &  &  & & & &  \\
    (3) Can't see  others' views$^{*}$ (PT) &  &  &  &  & & & &\\
     (4) Not sorry for others $^{*}$ (EC) &  &  & &  & & & &\\        
       (5)  Get involved in novels (FS)&  & & 0.8&  & & & &\\   
        (6)  Not-at-ease in emergencies (PD) & &  &  &  & & & & 0.7\\
         (7) Not caught-up in movies$^{*}$ (FS) &  & &  &  & & & &\\
          (8) Look at all sides in a fight (PT) &  & 0.9124  & 0.2444 &  & & & &\\
           (9) Feel protective of others  (EC) &  &  &  &  & &0.3 & &\\
           (10) Feel helpless when emotional (PD) &  &  &  &  & & & &\\
            (11) Imagine friend's perspective (PT) & &  &  &  & 0.8393& 0.824 & &\\
             (12) Don't get involved in books$^{*}$ (FS) & &  &  & & & & &\\
              (13) Remain calm if other's hurt $^{*}$ (PD)&  &  &  &  & & & &\\
               (14) Others' problems none mine$^{*}$ (EC) & &  &  & & & & &\\
                (15) If I'm right I won't argue$^{*}$ (PT)&  & &  &  & & & &\\
                 (16) Feel like movie character (FS) &  &  &  & & & & &\\
                  (17) Tense emotions scare me (PD) &  &  &  &  & & & &\\
                   (18) Don't feel pity for others $^{*}$ (EC)&  &  &  &  & & & &\\
                   (19) Effective in emergencies$^{*}$ (PD) &  &  &  &  & & & &\\
           (20) Touched by things I see  (EC)&  &  &  &  -0.3452& & & &\\
            (21) Two sides to every question  (PT)&  &  &  &  & & & &\\
             (22) Soft-hearted person (EC)&  &  &  &  & & & &\\
              (23) Feel like leading character (FS) & &  &  &  & & & &\\
               (44) Lose control in emergencies (PD)&  &  &  &  & & & &\\
                (25) Put myself in others' shoes (PT) &  &  &  &  & & & &\\
                 (26) Image novels were about me (FS)&  &  &  &  & & & &\\
                  (27) Other's problems destroy me (PD)& &  &  &  & & & &\\
                   (28) Put myself in other's place (PT)&  & 0.42&  &  & & & &\\
    \hline
  \end{tabular}
\end{table*}

\subsection*{Game questionnaire}
The second questionnaire  is about a prisoner dilemma game.  Each of the 47 participants is asked to answer with a yes or no  to 4 questions (see Table \ref{table:questionaire0}), each related to the level of cooperation - cooperation (CC), cooperation - defection (CD),  defection - cooperation (DC), and defection-defection (DD). A virtual other participant is represented in each interaction leading to 94 decision-makers in the whole process. The set of choices of each participant is $\{C,D\}$ where $D$ is also referred to $N$ for non-cooperation. 

\begin{table}[htb]
\centering
\begin{tabular}{cc|c|c|c|c|l}
\cline{3-4}
& & \multicolumn{2}{ c| }{Player I} \\ \cline{3-4}
& & Cooperate & Defect  \\ \cline{1-4}
\multicolumn{1}{ |c  }{\multirow{2}{*}{Player II} } &
\multicolumn{1}{ |c| }{Cooperate} & $(A,A)$ & $ (B,C)  $      \\ \cline{2-4}
\multicolumn{1}{ |c  }{}                        &
\multicolumn{1}{ |c| }{Defect } & $(C,B)  $ & $(D,D)$    \\
 \cline{1-4}
\end{tabular}
\caption[]{ Payoff matrix for standard prisoner's dilemma (without empathy consideration). The following inequalities must hold: $A >D>B>C$  \cite{b12}.}
\label{table:questionaire0}
\end{table}

\subsection*{Data collection}


Regarding the approach to the test, the whole population had a complete comprehension and adherence to the tasks. Only 2 questions have been left out  by a participant in the IRI test.  
In total we have a 99,63\% of responsiveness in all the questions. In the next section, we analyze the results of the second questionnaire and study the impact of the four IRI scales on the decision making of the population.

\section{Method and Analysis}

The analysis is divided into three steps.  In the first step,  the population is classified based on the result of the IRI scale. In the second step we analyze the result  of the cooperation.  Lastly, the level of cooperation is studied on the basis of  classification  of the population in the IRI scale. 

\subsection*{IRI scale and population classification}

The first step of the analysis  concerns the results of women and men population respectively at the IRI scale.  We depict the characteristic of each individual who participated to the test in table \ref{table:IRIdistra}. Table \ref{table:IRIdistr2b} represents the number of people belonging to each sub-scale and those who do not. 

\begin{table}[htb]
\centering
\begin{tabular}{lll}
{\bf Scale Type} & {\bf Women ID} & {\bf Men ID}  \\ \hline
PT  & 1,3, 4,10, 16,19,26,27 & 12 \\ \hline
EC     & 15 &  -    \\ \hline
FS     & 11,25   &   8 \\ \hline
PD     & -  &   18  \\ \hline

&& \\ \hline

PT + EC & 7,8,20& -  \\ \hline

PT + PD &  -   &  4,15 \\ \hline

PT + FS &  6,12,24 &  - \\ \hline

EC + FS & 17& 14  \\ \hline

EC + PD & 2& -  \\ \hline

&& \\ \hline

PT + EC + FS & 9,18,21& 5,19  \\ \hline

PT + EC + PD & 23& 6,11  \\ \hline

PT + FS + PD & - & 17  \\ \hline

EC + FS  + PD& 5& -  \\ \hline

&& \\ \hline

PT + EC + FS + PD    & 13,14,22 &   9  \\ \hline

&& \\ \hline

None of the scale  & 28   &   1,2,3,7,10,13,16  \\ \hline
\end{tabular}
\caption[]{ IRI scale and participant  identification}
\label{table:IRIdistra}
\end{table}

\begin{table}[htb]
\centering
\begin{tabular}{lllll}
{\bf Scale Type} & {\bf Women} & {\bf Men}  & {\bf  Total } &  {\bf Freq} \\ \hline

PT  & 8 & 1 & 9   & 19,14\% \\ \hline
EC     & 1 &  -    &1  &2,12 \%\\ \hline
FS     & 2   &   1 &3  & 6,38\% \\ \hline
PD     & -  &   1 &1 &2,12 \%\\ \hline

&& \\ \hline

PT + EC & 3& - &3 &6,38\% \\ \hline 

PT + PD &  -   &  2 &2 &4,25\% \\ \hline 

PT + FS &  3 &  - &3  &6,38\% \\ \hline 

EC + FS & 1& 1  &2 &4,25\% \\ \hline 

EC + PD & 1& -&1  &2,12 \%\\ \hline

&& \\ \hline

PT + EC + FS & 3& 2  &5 &10,63\% \\ \hline

PT + EC + PD & 1 & 2 &3 & 6,38\% \\ \hline

PT + FS + PD & - & 1 &1  &2,12 \%\\ \hline

EC + FS  + PD& 1& - &1  &2,12 \%\\ \hline

&& \\ \hline

PT + EC + FS + PD    & 3 & 1 &4 & 8,51\%  \\ \hline

&& \\ \hline

None of the scale  & 1 &   7  &8 & 17,02\% \\ \hline

  &    &   &  \\ \hline
Participants  &   28 &19    &  47  \\ \hline
\end{tabular}
\caption[]{ IRI scale and population distribution}
\label{table:IRIdistr2b}
\end{table}

The classification of the population based on different empathy subscales is presented in  Table  \ref{table:IRIdistr2b}.  The result shows that 14 people belong to a pure IRI scale while 25 people has a mixed IRI characteristics and  8 people do not belong to any IRI scale.  In the next section we study the level of cooperation based on the classification of the population in the IRI scale.

\subsection*{Cooperation study: Prisoner Dilemma}
The analysis is based on the IRI test results and on the prisoner's dilemma  test results.  The result of the prisoner's dilemma suggested that the 35,71 \% of the women  population and  the 36,84 \% of  men population have fully confessed.  The results  are depicted in Table \ref{table:fem1}, \ref{table:mal1} and \ref{table:pop1}. Notice that 53,57 \% of the women population  and 31,57 \% of the men population have partially confessed. When considering the whole population, 23,40\% of them has partially confessed. Hence, it is  necessary to classified them looking at the cooperation level within the population of those who partially confessed.

\begin{table}[htb]
\centering
\begin{tabular}{|c|c|c|c|}
\hline
\multicolumn{4}{|c|}{ { \bf Cooperation results (Women) } } \\ \hline
Decision & positively &  partially  &  deny  \\ \hline
Result&  10 out of 28  &   15 out of 28  &  3 out of 28 \\ \hline
Frequency & 35,71\%  & 53,57\%  & 10,71\%\\ \hline
\end{tabular}
\caption[]{ Cooperation results: Women}
\label{table:fem1}
\end{table}

\begin{table}[htb]
\centering
\begin{tabular}{|c|c|c|c|}
\hline
\multicolumn{4}{|c|}{ { \bf Cooperation results (Men) } } \\ \hline
Decision & positively &  partially  &  deny  \\ \hline
Result&   7 out of 19  &  6 out of 19 &  6  out of 19 \\ \hline
Frequency & 36,84 \%  & 31,57\%  & 31,57\%\\ \hline
\end{tabular}
\caption[]{ Cooperation results: Men}
\label{table:mal1}
\end{table}

\begin{table}[htb]
\centering
\begin{tabular}{|c|c|c|c|}
\hline
\multicolumn{4}{|c|}{ { \bf Cooperation results (Entire population) } } \\ \hline
Decision & positively &  partially  &  deny  \\ \hline
Result&  17  out of 47  & 21 out of 47 &   9  out of 47 \\ \hline
Frequency &   36,17 \%  & 44,68\%  &  19,14\% \\ \hline
\end{tabular}
\caption[]{ Cooperation results: Entire Population}
\label{table:pop1}
\end{table}

A more refined version of cooperation among  the women population who partially confessed (15 out of 28) is given in Table \ref{table:par 1}. We want to compute the level of cooperation within that population and hence we care about all the answer of  the participants at the questionnaire. Then we derive the  level of cooperation in that population by computing the marginal probability of confess in the population. More precisely, we consider  two random variables  $X =\{ c_1, d_1\}$ and   $Y= \{ c_2, d_2\}$  for player 1 and player 2 respectively where $c_i$ stands for cooperation of player $i$ and $d_i$ defection. We then compute the marginal probability of  cooperation of the player 1  given that the player 2 can confess or defect. $$p(c_1) = \sum_{y \in   \{ c_2, d_2\}}  p(X  = c_1, Y = y).$$ We use the sample statistics to compute the probability of player 1 to cooperate through the number of occurrences of $c_1$.
The result of the marginal probability of cooperation is equals to $0,51$. Hence in average, 7 people out the 15  can be classified as positively confessed. 
\begin{table}[htb]
\begin{tabular}{cc|c|c|c|c|l}
\cline{3-4}
& & \multicolumn{2}{ c| }{Player I} \\ \cline{3-4}
& & Cooperate & Defect  \\ \cline{1-4}
\multicolumn{1}{ |c  }{\multirow{2}{*}{Player II} } &
\multicolumn{1}{ |c| }{Cooperate} & 10 & $ 1  $      \\ \cline{2-4}
\multicolumn{1}{ |c  }{}                        &
\multicolumn{1}{ |c| }{Defect } & $5  $ & 13   \\
 \cline{1-4}
\end{tabular}
\caption[]{ Women population: Partially confess (decision making). $p(c_1) =  0.51$ }
\label{table:par 1}
\end{table}

In the case of the men, a more refined version of cooperator among those who partially confess (6 out of 19) is given in Table \ref{table:par 2}.  The marginal probability of cooperation is equal to $p(c_1) = 0,46$. Hence in average, 2 people out the 6  can be classified as positively confessed. 

\begin{table}[htb]
\begin{tabular}{cc|c|c|c|c|l}
\cline{3-4}
& & \multicolumn{2}{ c| }{Player I} \\ \cline{3-4}
& & Cooperate & Defect  \\ \cline{1-4}
\multicolumn{1}{ |c  }{\multirow{2}{*}{Player II} } &
\multicolumn{1}{ |c| }{Cooperate} & 5 & $ 2  $      \\ \cline{2-4}
\multicolumn{1}{ |c  }{}                        &
\multicolumn{1}{ |c| }{Defect } & $1  $ & 5   \\
 \cline{1-4}
\end{tabular}
\caption[]{ Men: Partially confess (decision making).  $p(c_1) =  0.46.$ }
\label{table:par 2}
\end{table}

When considering the whole population, the fraction of people who partially confess is  21/47 and  the marginal probability of cooperation  is equals to $p(c_1) = 0,5$. Hence in average, 10 people out of the 21 can be classified as positively confessed.


\begin{table}[htb]
\begin{tabular}{cc|c|c|c|c|l}
\cline{3-4}
& & \multicolumn{2}{ c| }{Player I} \\ \cline{3-4}
& & Cooperate & Defect  \\ \cline{1-4}
\multicolumn{1}{ |c  }{\multirow{2}{*}{Player II} } &
\multicolumn{1}{ |c| }{Cooperate} & 15 & $ 3  $      \\ \cline{2-4}
\multicolumn{1}{ |c  }{}                        &
\multicolumn{1}{ |c| }{Defect } & $6  $ & 18   \\
 \cline{1-4}
\end{tabular}
\caption[]{ Population: Partially confess (decision making).  $p(c_1) = 0.5$ }
\label{table:par 3}
\end{table}


\subsection*{Cooperation vs IRI scale}
In this subsection, we are interested in computing the level of cooperation in each IRI 'pure' subscale. For this aim, we consider the subpopulation belonging to each scale.  We then  use the answer of cooperation in the prisoner dilemma game for computing the probability of cooperation. \\ \\ 

{ \bf \color{black} PT vs Cooperation}

\begin{minipage}{0.4\textwidth}
\centering

 \begin{tabular}{|c|c|c|c|c|c|}
\hline
 \multicolumn{6}{|c|}{  \bf  \color{black}  Women  } \\ \hline
{\color{black}  Coop\textbackslash  PT  } & A  &  B & C &  D & E  \\ \hline
p(c)  &0 \% &0.5 \% & 0.75\% & 0.66\% &100\%  \\ \hline
\end{tabular}

\vspace{0.3 cm}

\centering
 \begin{tabular}{|c|c|c|c|c|c|}
\hline
 \multicolumn{6}{|c|}{ \bf  \color{black}  Men } \\ \hline
{\color{black}  Cooperation\textbackslash  PT  } & A  &  B & C &  D & E  \\ \hline
p(c)  &0 \% &0 \% & 0\% & 66,66\% &0\%  \\ \hline
\end{tabular}

\vspace{0.3 cm}

 \begin{tabular}{|c|c|c|c|c|c|}
\hline
 \multicolumn{6}{|c|}{ \bf  \color{black}  Women  + Men  } \\ \hline
{\color{black}  Coop\textbackslash  PT  } & A  &  B & C &  D & E  \\ \hline
p(c)  &0 \% &0,5 \% & 75\% & 66\% &100\%  \\ \hline
\end{tabular}
\end{minipage}  

The Pearson correlation coefficient between the level of cooperation and the PT scale is  $r_{Women}$$ = 0.7797$ $(p < .01)$ for Women, $r_{men}$ = 1 $(p < .01)$ for men and $r_{population}$ $= 0.6347$ $(p < .01)$   for the population belonging to the PT scale. The interpretation is that there is a positive correlation for Women. The fact that only one man was PT and had positively cooperate leads to a strong positive correlation. The overall population of PT positively cooperates.

\vspace{ 0.3cm}

{ \bf \color{black} PD vs Cooperation}

\begin{minipage}{0.4\textwidth}
\centering
 \begin{tabular}{|c|c|c|c|c|c|}
\hline
 \multicolumn{6}{|c|}{  \bf  \color{black}  Men (18)  } \\ \hline
{\color{black}  Cooperation\textbackslash  PD  } & A  &  B & C &  D & E  \\ \hline
p(c) & 0 \%&0 \% &   { \color{black}  0 \% } & 0\% &0\%  \\ \hline
\end{tabular}
\end{minipage}

There is only one  man who is  PD in the IRI scale and the probability of his level of cooperation was zero since he only denied.

\vspace{ 0.3 cm}

{ \bf \color{black} EC vs Cooperation: Women}

\begin{minipage}{0.4\textwidth}
\centering
 \begin{tabular}{|c|c|c|c|c|c|}
\hline
 \multicolumn{6}{|c|}{  \bf  \color{black}  Women (15)  } \\ \hline
{\color{black}  Cooperation\textbackslash  EC  } & A  &  B & C &  D & E  \\ \hline
p(c) &  0 \% &0 \% & { \color{black}  100 \% }  & 0\% &0\%  \\ \hline
\end{tabular}
\end{minipage}

There is only one  woman EC in the IRI scale and the probability of his level of cooperation is 1 since she positively confessed. Therefore the Pearson correlation coefficient is one.

\vspace{ 0.3 cm}

{ \bf \color{black} FS vs Cooperation}

\begin{minipage}{0.4\textwidth}
\centering

 \begin{tabular}{|c|c|c|c|c|c|}
\hline
 \multicolumn{6}{|c|}{  \bf  \color{black}  Women (11,25)  } \\ \hline
{\color{black}  Coop\textbackslash  FS  } & A  &  B & C &  D & E  \\ \hline
p(c) &0 \% &0 \% & 0 \% &  { \color{black} 100\% } &0\%  \\ \hline
\end{tabular}

\vspace{0.3 cm}

\centering
 \begin{tabular}{|c|c|c|c|c|c|}
\hline
 \multicolumn{6}{|c|}{ \bf  \color{black}  Men (8)  } \\ \hline
{\color{black}  Coop\textbackslash  FS  } & A  &  B & C &  D & E  \\ \hline
p(c) &0 \% &0 \% & 0\% & 0\% &  { \color{black}  66,66\% }  \\ \hline
\end{tabular}

\vspace{0.3 cm}

 \begin{tabular}{|c|c|c|c|c|c|}
\hline
 \multicolumn{6}{|c|}{ \bf  \color{black}  Women  + Men  } \\ \hline
{\color{black}  Coop.\textbackslash  FS  } & A  &  B & C &  D & E  \\ \hline
p(c) &0 \% &0 \% & 0\% & { \color{black} 100\% } &{ \color{black} 66,66\% }  \\ \hline
\end{tabular}

\end{minipage}

The Pearson correlation coefficient between the level of cooperation and the FS scale is  $r_{Women}$$ = 1$ $(p < .01)$ for Women,  $r_{men}$ = 1 $(p < .01)$ for men and $r_{population}$ $=  0.9891$ $(p < .01)$   for the population belonging to the FS scale.  The strong positive correlation between FS and cooperation is due to the fact that  two people had positively confessed and one had partially confessed.


\pgfplotsset{
  compat=newest,
  xlabel near ticks,
  ylabel near ticks
}
\begin{tikzpicture}[font=\small]
    \begin{axis}[
      ybar interval=0.3,
      bar width=2pt,
      grid=major,
      xlabel={Empathy Scale Quality: Perspective Taking  (PT)},
      ylabel={Total Score of Answers},
      ymin=0,
      ytick=data,
      xtick=data,
      axis x line=bottom,
      axis y line=left,
      enlarge x limits=0.1,
      symbolic x coords={awful,bad,average,good,excellent,ideal},
      xticklabel style={anchor=base,yshift=-0.5\baselineskip},
    ]
           \addplot[fill=yellow] coordinates {
               (awful,3) (bad,13)  (average,14)   (good,16)  (excellent,9) (ideal, 20)
  }; \addplot[fill=white] coordinates {
      (awful,1)  (bad,1)  (average,2)  (good,3)   (excellent,0) (ideal, 20)      
      };

\legend{ Women, Men}
    \end{axis}
  \end{tikzpicture}

\pgfplotsset{
  compat=newest,
  xlabel near ticks,
  ylabel near ticks
}
\begin{tikzpicture}[font=\small]
    \begin{axis}[
      ybar interval=0.3,
      bar width=2pt,
      grid=major,
      xlabel={Empathy Scale Quality: Personal Distress  (PD)},
      ylabel={Total Score of Answers},
      ymin=0,
      ytick=data,
      xtick=data,
      axis x line=bottom,
      axis y line=left,
      enlarge x limits=0.1,
      symbolic x coords={awful,bad,average,good,excellent,ideal},
      xticklabel style={anchor=base,yshift=-0.5\baselineskip},
    ]
           \addplot[fill=yellow] coordinates {
               (awful,0) (bad,0)  (average,0)   (good,0)  (excellent,0) (ideal, 10)
  }; \addplot[fill=white] coordinates {
      (awful,0)  (bad,2)  (average,4)  (good,1)   (excellent,0) (ideal, 10)      
      };

\legend{ Women, Men}
    \end{axis}
  \end{tikzpicture}

\pgfplotsset{
  compat=newest,
  xlabel near ticks,
  ylabel near ticks
}
\begin{tikzpicture}[font=\small]
    \begin{axis}[
      ybar interval=0.3,
      bar width=2pt,
      grid=major,
      xlabel={Empathy Scale Quality: Empathic Concern (EC)},
      ylabel={Total Score of Answers},
      ymin=0,
      ytick=data,
      xtick=data,
      axis x line=bottom,
      axis y line=left,
      enlarge x limits=0.1,
      symbolic x coords={awful,bad,average,good,excellent,ideal},
      xticklabel style={anchor=base,yshift=-0.5\baselineskip},
    ]
           \addplot[fill=yellow] coordinates {
               (awful,2) (bad,0)  (average,2)   (good,3)  (excellent,0) (ideal, 10)
  }; \addplot[fill=white] coordinates {
      (awful,0)
       (bad,0)
        (average,0)
         (good,0)
        (excellent,0) (ideal, 0)      
      };

\legend{ Women, Men}
    \end{axis}
  \end{tikzpicture}

\pgfplotsset{
  compat=newest,
  xlabel near ticks,
  ylabel near ticks
}
\begin{tikzpicture}[font=\small]
    \begin{axis}[
      ybar interval=0.3,
      bar width=2pt,
      grid=major,
      xlabel={Empathy Scale Quality: Fantasy Scale (FS)},
      ylabel={Total Score of Answers},
      ymin=0,
      ytick=data,
      xtick=data,
      axis x line=bottom,
      axis y line=left,
      enlarge x limits=0.1,
      symbolic x coords={awful,bad,average,good,excellent,ideal},
      xticklabel style={anchor=base,yshift=-0.5\baselineskip},
    ]
           \addplot[fill=yellow] coordinates {
               (awful,1) (bad,3)  (average,3)   (good,6)  (excellent,2) (ideal, 10)
  }; \addplot[fill=white] coordinates {
      (awful,1)  (bad,1)  (average,1)    (good,1)  (excellent,3) (ideal, 10)      
      };

\legend{ Women, Men}
    \end{axis}
  \end{tikzpicture}

\subsection*{Cooperation vs IRI mixed scale}  \label{newmodel1}

In this section we analyze the correlation between the  mixed scale of length two and we  also compute the level of cooperation in each sub-population.

\begin{itemize}
\item  Case {\bf PT + EC}: the sub-population  is composed of only women.  The Pearson correlation coefficient between PT and EC is   $r_{Women} =$ $r_{population}$ $ = 0,8108$ $(p < .01).$ The probability of cooperation was p(c) =  0,5.

\item  Case {\bf PT + FS}: the sub-population  is composed of only women.  The Pearson correlation coefficient between PT and FS is   $r_{Women} =$ $r_{population}$ $ = 0,9382$ $(p < .01).$ The probability of cooperation was p(c) =  0,62.

\item  Case {\bf EC + FS}: the sub-population  is composed of  women and men.  The Pearson correlation coefficient between EC and FS  is   $r_{Women} =$  $ = 0,7845$ $(p < .01)$,  $r_{Men} =$  $ = 0,8709$ $(p < .01),$ $r_{population}$ $ = 0,7148$ $(p < .01)$  for women, men and the global population respectively. The probability of cooperation was p(c) =  0,75.

\item  Case {\bf PT + PD}: the sub-population  is composed of only men.  The Pearson correlation coefficient between PT and PD is   $r_{Men} =$ $r_{population}$ $ = 0,2796$ $(p < .01).$ The probability of cooperation was p(c) =  0,6.

\item  Case {\bf EC + PD }: the sub-population  is composed of only women.  The Pearson correlation coefficient between PT and FS is   $r_{Women} =$ $r_{population}$ $ = -0,3462$ $(p < .01).$ The probability of cooperation was p(c) =  0,5.

\end{itemize}

\begin{table}[htbp]
\begin{center}
\begin{tabular}{|c|c|c|c|c|} 
 \hline
Pearson correlation  & PT  & EC  & FS  & PD \\ \hline
PT & -& { \bf 0,81}& { \bf0,9382} & 0,2796  \\ \hline
EC & -&- & { \bf0,8709}& -0,3462  \\ \hline
FS & - &- & -& - \\ \hline
PD & -&- &- &-  \\ \hline
\end{tabular}
\end{center}
\caption{subscale correlation}
\label{ms}
\end{table}%

The result of level of cooperation corresponding to the mixed IRI scale of Table  \ref{ms}  is  given in Table  \ref{mscoop}. We can observe that a high level of cooperation associated to a high correlation coefficient correspond to  the Empathy-Altruism behaviour (namely  PT + FS  and EC + FS ).  A high level of cooperation associated to a low correlation coefficient corresponds to Empathy-Spitefulness (namely PT + PD, EC + PD).

\begin{table}[htbp]
\begin{center}
\begin{tabular}{|c|c|} 
 \hline
 & Cooperation level \\ \hline
PT  + EC & 50\%  \\ \hline
PT + FS &  62,5\%   \\ \hline
PT + PD & 66,66 \% \\ \hline
EC + FS & 75\% \\ \hline
EC + PD & 50\% \\ \hline
\end{tabular}
\end{center}
\caption{Level of cooperation at mixed scales}
\label{mscoop}
\end{table}%

{\color{black}

\subsection*{The effect of empathy on decisions}  \label{empathyDecisions}  

In this section we study the effect of individual scales on the degree of cooperation.  For this aim we will compare the result of a pure scale with the group of individual who do not belong to any IRI sub-scale. The motivation of this placebo test  is that people who do not belongs to any scale can be a valuable sample for assessing the impact of  an IRI scale like PT, EC and FS  on user's decision making.

Since our dataset is not too large, we rely on the nonparametric linear regression using the Theil's method  \cite{Theil1950a,Theil1950b,Theil1950c} for computing the  slope median value, given a dependent variable  set $\{y\}$ and an  independent variable $\{x\}$.

The dataset of the independent variable  $\{x\}$ is represented by an IRI scale and it  is obtained in the following way:  (i) we first select the scale we want to study ( let say PT scale).  The cardinality of the dataset is  then given by  the number of people belonging to that scale (see Table \ref{table:IRIdistra}). Next the value $x_i$ associate to the individual $i$ is computed by choosing the value (let say A)  with the highest choice within the questionnaire
$\{ A, B, C, D,E\}$. Based on the choice's result, an integer value is assigned to   $\{x_i\}$  following  $\{  A= 1, B = 2, C= 3, D=4, E = 5 \} $.

The dataset of the dependable variable $\{y\}$ is represented by the result of the cooperation.  An individual $i$ will be assigned a value $y_i = 1$ if he has fully cooperated,  $y_i = 0.5$ if he has partially cooperated and $y_i = 0$ if he has denies.

Based on the  above definition we can now  study the effect of  the PT scale on the level of cooperation. There are 9 individual belonging to the pure PT scale (see Table \ref{table:PTlr} ).
\begin{table} [h]
\centering
\begin{tabular}{|c|c|c|c|c|c|c|c|c|c|}
\hline
 \multicolumn{10}{|c|}{ { \bf  PT level: pure PT individual  }} \\ \hline
{\color{black}  x} & 2 & 3 & 3 & 4	& 4 & 4 &	4 & 4 & 5  \\ \hline
{\color{black}  y } & 0.5 &	0.5 & 	1 &	0 &	0.5	& 1 &	1 &	1 &	1 \\ \hline
 \multicolumn{10}{|c|}{ { \bf  Median slope: $\beta_{PT} =  \bf{0.2515}$   }} \\ \hline
\end{tabular}
\caption[]{  { \bf PT: nonparametric linear regression dataset}}
\label{table:PTlr}
\end{table}

The result of the Theil's slope median is given by  $\beta_{PT} =  \bf{0.2515}$,  where   $\beta_{PT} $ is the median slope value of the set 

 $\{ -1.0010,   -0.5010,   -0.4980,   -0.2500,    0.001,    0.001,    0.001, \\  0.001,   0.001,    0.002,    0.002,    0.002,    0.003,   0.003, \\   
 0.1680,     0.2502,     0.2506,     0.2510,    {  \color{black} \bf{0.2515} },     0.4990,     0.4995,  \\  0.4995,    0.5000,    0.5025,   1,    1,    1,    1.004, \\ 
 167,  250,  250.75,  334,  499,  499, \\ 500.5,  502 \}$

 For  the placebo test, the dataset for the individual belonging to ``None of the scale'' is given by the Table \ref{table:nonPTlr}

 \begin{table} [h]
\centering
\begin{tabular}{|c|c|c|c|c|c|c|c|c|}
\hline
 \multicolumn{9}{|c|}{ { \bf PT level: None of the scale  individual }} \\ \hline
{\color{black}  x} & 1	&2	&2	&2	&2	&2	&4&	4 \\ \hline
{\color{black}  y } & 0& 	0& 	0& 	0.5& 	0.5& 	1& 	0.5	& 1\\ \hline
 \multicolumn{9}{|c|}{ { \bf  Median slope: $\beta_{non \,\, of  \,\, the  \,\, scale} =  \bf{0.4995}$   }} \\ \hline
\end{tabular}
\caption[]{  { \bf PT level: nonparametric linear regression dataset (none of the scale)}}
\label{table:nonPTlr}
\end{table}

The result of the Theil's slope median is given by  $\beta_{\mbox{non of  the scale}} =  \bf{0.4995}$,  where   $\beta_{\mbox{non of   the  scale}} $ is the median slope value of the set 

 $\{
 -0.2495,    0.0005,    0.0005,    0.001 ,   0.001,    0.002 ,   0.1673, \\
 0.2501,    0.2503,    0.2505,    0.2506,    0.3336,    0.499,     {  \color{black} \bf{ 0.4995 } },  \\
 0.4995 ,   0.4998,    0.996 ,   1,    1,  166.6667,   249.5 ,  \\
 249.5,   249.75,   250,   332.6667 ,  498 ,   499 ,  \\   499
   \}$ 
  
  Result interpretation: the angular coefficient $\beta$ measures the effect of the independent variable $x$ on the dependent variable $y$. The more is the value of the angular coefficient, the better is the effect of the independent  variable $x$ on the variable  $y$.  The results we get are  $\beta_{PT} =  0.2515$ and $\beta_{non \,\, of  \,\, the  \,\, scale} =  0.4995$ and  $\beta_{PT}  <  \beta_{non \,\, of  \,\, the  \,\, scale}$. Since we are studying the effect of the PT scale on the level of cooperation and based on the following result, we cannot concludes that the only factor which increases  the level of cooperation is  given by the PT component.

  Similarly,  we are interested in studying the influence of Fantasy scale component on the level of cooperation and we follow and apply  the approach mentioned above on the FS component. 
  
   The result of the Theil's slope median is given by  $\beta_{FS} =  \bf{0.5000  }$,  where   $\beta_{FS} $ is the median slope value (see Table \ref{table:FSlr} ) of the set  $\{ -0.0010,      {  \color{black} \bf{ 0.5000 } },  501.0000  \}$ 
 \begin{table} [h]
\centering
\begin{tabular}{|c|c|c|c|   }
\hline
 \multicolumn{4}{|c|}{ { \bf  FS level: pure FS individual  }} \\ \hline
{\color{black}  x} & 4&	4	&5 \\ \hline
{\color{black}  y } & 0.5&	1&	1 \\ \hline
 \multicolumn{4}{|c|}{   { \bf  Median slope: $\beta_{FS} =  \bf{0.5} $     }   } \\ \hline
\end{tabular}
\caption[]{  { \bf FS: nonparametric linear regression dataset}}
\label{table:FSlr}
\end{table}

In the case of  individual belonging to ``None of the scale'', the result of the Theil's slope median is given by  $ \beta_{non \,\, of  \,\, the  \,\, scale} = 0.5000 $    where,    $\beta_{non \,\, of  \,\, the  \,\, scale} $ is the median slope  (see Table \ref{table:nonFSlr} ) value of the set   

$\{    -0.5005,   -0.499,    0.001,    0.0010,    0.001,    0.002,   0.002, \\ 
    0.2504,    0.2508 ,   0.4995 ,   0.4995 ,   0.5 ,   0.5003 ,   {  \color{black} \bf{ 0.5005  }},  \\ 
    0.5010,    0.5015,    0.9980,    0.9980,    0.999,    \\ 1,    1, 
  167,  250,  250,  498,  499,  499,   500
   \}$

 \begin{table} [h]
\centering
\begin{tabular}{|c|c|c|c|c|c|c|c|c|}
\hline
 \multicolumn{9}{|c|}{ { \bf FS level: Non of the scale  individual }} \\ \hline
{\color{black}  x} & 1	&1	&1	&1&	2	&2&	2	&3\\ \hline
{\color{black}  y } &   0 	&0	&0.5&	0.5&	0	&0.5&	1&	1  \\ \hline
 \multicolumn{9}{|c|}{ { \bf  Median slope: $\beta_{non \,\, of  \,\, the  \,\, scale} =  \bf{0.5005}$   }} \\ \hline
\end{tabular}
\caption[]{  { \bf FS level: nonparametric linear regression dataset (none of the scale)}}
\label{table:nonFSlr}
\end{table}
The dataset of the component EC and PD is to small to perform the nonparametric linear regression on it.

}  


\subsection*{Explanation} \label{newmodel}

\subsubsection*{Game without empathy}
We consider the one-shot game given by Table \ref{table:noEmpathy}.
{\color{black} Pareto efficiency, or Pareto optimality, is an action profile  in which it is not possible to make any one player better off without making at least one player  worse off. A Nash equilibrium is a situation in which no player can improve her payoff by unilateral deviation.
The action profile $(D,D)$ is the unique Nash equilibrium, and $D$ is a dominant strategy choice for each player. But $(C,C)$ Pareto-dominates $(D,D).$ The three choice 
pairs $(C,C)$, $(C,D),$ and $(D,C)$ are all Pareto optimal, but $(C,C)$ is the most socially efficient choice pair.}

\begin{table}[htb]
\centering
\begin{tabular}{cc|c|c|c|c|l}
\cline{3-4}
& & \multicolumn{2}{ c| }{Player I} \\ \cline{3-4}
& & Cooperate & Defect  \\ \cline{1-4}
\multicolumn{1}{ |c  }{\multirow{2}{*}{Player II} } &
\multicolumn{1}{ |c| }{Cooperate} & (-6,-6) & $ (-120,0)  $      \\ \cline{2-4}
\multicolumn{1}{ |c  }{}                        &
\multicolumn{1}{ |c| }{Defect } & $(0,-120)  $ & (-72,-72)    \\
 \cline{1-4}
\end{tabular}
\caption[]{ Payoff matrix of prisoner's dilemma questionnaire }
\label{table:noEmpathy}
\end{table}

The classical game model fails to explain the experimental observation:

In the classical prisoner dilemma (without empathy consideration),  the strategy which consists to defect is a dominated strategy.  Hence it is expected that - in the rational case -  all the people decides to defect in the questionnaire of the first experiment (see Table \ref{table:noEmpathy}). But this was not the case since   19,14\% of the population defected (see Table \ref{table:pop1}).  This result can be due to some psychological aspect of the human behavior when taking part of the game.  The idea is to modify the classical payoff and integrate the empathy in the preferences of the players. This leads to an empathetic payoff as explained below.


\subsubsection*{Game with empathy consideration}
As observed from the data, a significant level of cooperation appears in the experimental game. This  suggests  a new  modelling and design of the classical game and better understanding the behavior of the participants.  We propose a a new payoff matrix that takes into consideration the effect of empathy in the outcome. Denote by $\lambda_{12}$ the degree of empathy of  the prisoner $1$ has over the prisoner $2$ and by $\lambda_{21}$  the vice versa. The  payoff  of the classical prisoner dilemma game (see Table \ref{table:noEmpathy}) changes and now depends on the level of empathy  $\lambda_{12}$ and $\lambda_{21}$ of the prisoners (see Table \ref{table:Empathyconsideration}). 
Now we are interested in finding all the possible equilibria of the new game based on the value of   $\lambda_{12}$ and $\lambda_{21}$. 

\begin{table}[htb]
\footnotesize 
\centering
\begin{tabular}{cc|c|c|c|c|l}
\cline{3-4}
& & \multicolumn{2}{ c| }{P  I} \\ \cline{3-4}
& & C & D  \\ \cline{1-4}
\multicolumn{1}{ |c  }{\multirow{2}{*}{P II} } &
\multicolumn{1}{ |c| }{C} & $(-6-6 \lambda_{12}, -6 -6\lambda_{21})$  & $ (-120, -120\lambda_{21})  $      \\ \cline{2-4}
\multicolumn{1}{ |c  }{}                        &
\multicolumn{1}{ |c| }{D} & $(-120\lambda_{12},-120)$ & $(-72-72\lambda_{12}, -72-72\lambda_{21})$    \\
 \cline{1-4}
\end{tabular}
\caption[]{ Payoff matrix of the prisoner dilemma with empathy consideration.}
\label{table:Empathyconsideration} 
\end{table}


{ Equilibrium analysis of Table \ref{table:Empathyconsideration} }
\begin{itemize}
\item  CC  is an equilibrium if  $\lambda_{12} \geq   \frac{6}{114};$  $\lambda_{21} \geq   \frac{6}{114}$
\item CN is an equilibrium if $\lambda_{12} \geq   \frac{2}{3};$  $\lambda_{21} \leq   \frac{6}{114}$
\item  NC is an equilibrium if $\lambda_{12} \leq   \frac{6}{114};$  $\lambda_{21} \geq   \frac{2}{3}$
\item  NN is an equilibrium if $\lambda_{12} \leq   \frac{2}{3};$  $\lambda_{21} \leq   \frac{2}{3}$
\end{itemize} Since empathy can be positive, negative or null,  we analyze the outcome of the game with different signs of the  parameters   $\lambda_{12}$ and $\lambda_{21}$. \\ \\ { \bf Analysis 1: $\lambda_{12}, \lambda_{21}  \geq 0$ }
 we now consider  $\lambda_{12}$ and $\lambda_{21}$  as two random variables with a distribution across the population. We generalized the outcome of the game based on their values. 
 \begin{itemize}
\item  {\bf case 1:}  {\bf (medium-medium)}: if   $\lambda_{12},\lambda_{21} $ $ \in$ $\left[  \frac{6}{114},  \frac{2}{3} \right] $   then we have  3 equilibria: CC, NN, and the mixed equilibria  $p_2C + (1-p_2)N$ or  $p_1C + (1-p_1)N$ with  $p_1 =  \frac{8 - 12\lambda_{12} }{ 7 (1 + \lambda_{12})}$ and  $p_2 =  \frac{8 - 12\lambda_{21} }{ 7 (1 + \lambda_{21})}$.
\item   {\bf case 2 (high - high):} if $\lambda_{12}, \lambda_{21}$ $ \in$ $\left[  \frac{2}{3},1  \right] $ then CC is the unique equilibrium. 
\item    {\bf case  3$_{a}$ (high - low):} if $\lambda_{12}$ $ \in$ $\left[  \frac{2}{3},1  \right] $  and  $\lambda_{21} \in \left[ 0,  \frac{6}{114} \right] $   then CN is the unique equilibrium. 
\item     {\bf case 3$_{b}$ (low -  high):} if $\lambda_{12}$   $\in$  $\left[ 0,  \frac{6}{114}\right] $ and  $\lambda_{21} \in  \left[  \frac{2}{3},1  \right] $    then NC is the unique equilibrium.
\item   {\bf case 4$_{a}$ (medium -  low):} if $\lambda_{12}$   $\in$  $\left[  \frac{6}{114}, \frac{2}{3}  \right] $ and  $\lambda_{21} \in  \left[ 0,  \frac{6}{114}    \right] $    then NN is the unique equilibrium. 
\item  {\bf case 4$_{b}$ ( low - medium ):} if $\lambda_{12}$   $\in$ $  \left[ 0,  \frac{6}{114}    \right] $  and  $\lambda_{21} \in$   $\left[  \frac{6}{114}, \frac{2}{3}  \right] $    then NN is the unique equilibrium.
\item      {\bf case 5$_{a}$  (   $\lambda_{12}$ high   ):} if $\lambda_{12} > \frac{2}{3} $ then C is a dominating strategy (unconditional cooperation) for player 1.
\item      {\bf case 5$_{b}$ (   $\lambda_{21}$ high   ):} if $\lambda_{21} > \frac{2}{3} $ then C is a dominating strategy (unconditional cooperation). for player 2.
\item        {\bf case 6$_{a}$ (   $\lambda_{12}$ low  ):}    if $\lambda_{12}$   $\in$  $\left[ 0,  \frac{6}{114}\right] $  then N is a dominating strategy (unconditional non-cooperation) for player 1.
\item       {\bf case 6$_{b}$ (   $\lambda_{21}$ low   ):}  if $\lambda_{21}$   $\in$  $\left[ 0,  \frac{6}{114}\right] $ then N is a dominating strategy (unconditional non-cooperation) for player 2.
\end{itemize} 
  { \bf Analysis 2: $\lambda_{12}, \lambda_{21}  <  0$ }  
\begin{itemize}
\item     if   $\lambda_{12}, \lambda_{21}  <  0$  then NN is the unique equilibrium. 
\end{itemize}
 { \bf Analysis 3: $\lambda_{12}>0, \lambda_{21}  <  0$ } 
\begin{itemize}
\item  if  $\lambda_{21}  <  0$ then   N is the dominating strategy (unconditional non-cooperation)   for player 2.
\item    if $\lambda_{12} > \frac{2}{3} $ then CN is an equilibrium. 
\item    if $\lambda_{12} < \frac{2}{3} $ then NN is an equilibrium. 
\end{itemize}
{ \bf Analysis 4: $\lambda_{12}<0, \lambda_{21}  >  0$ } 
\begin{itemize}
\item  if  $\lambda_{12}  <  0$ then  N is a dominating strategy (unconditional non-cooperation)   for player 1.
\item    if $\lambda_{21} > \frac{2}{3} $ and $\lambda_{12}$   $\in$  $\left[ - \frac{6}{114}, 0  \right] $ then NC is an equilibrium. 
\item    if $\lambda_{21} < \frac{2}{3} $ then NN is an equilibrium. 
\end{itemize}

\begin{figure}[htb]
\includegraphics[scale=0.27]{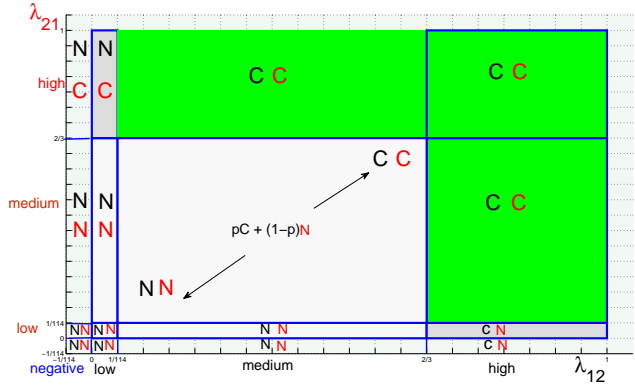}\\ \caption{Equilibrium of the game with Empathy consideration}
\label{fig:twoEmpathy}
\end{figure}

Figure \ref{fig:twoEmpathy}  summarizes the outcome of the two-player game.
 
The proposed empathetic payoff better captures the preferences of the players as a negligible proportion of cooperators are obtained analytically in the new game. Thus, if one quantity accurately the empathy's effect in the payoff, then the resulting game is more adapted to the results of experiment. By doing this iteratively over several experiments and model adjustment we obtain better game theoretic models for real life interaction. We believe that the generic approach developed here can be extended to other class of games as indicated in \cite{t1,mimo}.

 \def\angle{0}
\def\radius{8}
\def\cyclelist{{"green", "blue","orange","red", "yellow"}}
\newcount\cyclecount \cyclecount=-1
\newcount\ind \ind=-1


\begin{figure}[htb]
\begin{center}
\centering
\includegraphics[width=10cm,height=10cm]{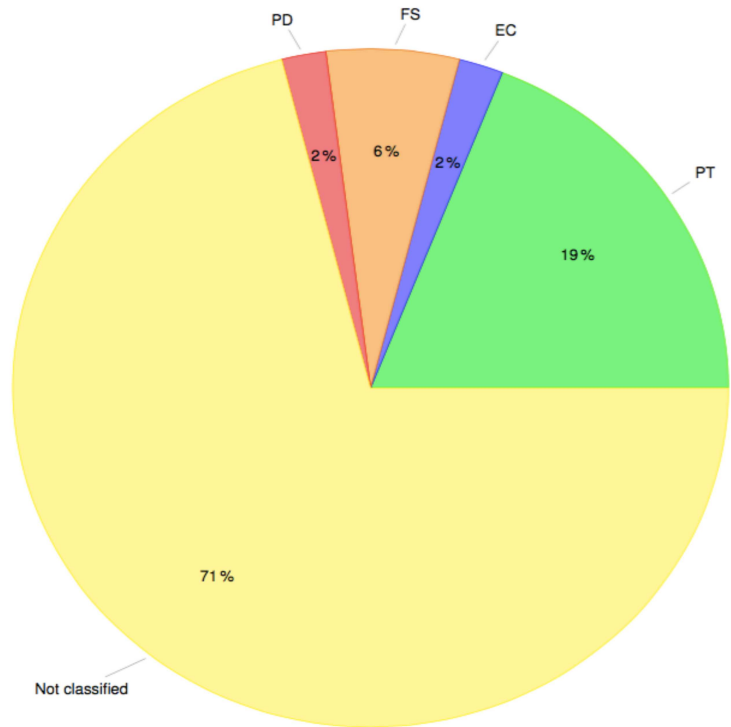}
\caption{   Empathy scale distribution across the population of participants   }
\label{fig2:repartition}
\end{center}
\end{figure}

\section{Conclusion}

We have proposed a basic experiment  on the role of empathy in one-shot Prisoner Dilemma. We  analyzed multidimensional components of empathy using IRI scale. The experiment on the field conducted at NYUAD, Learning and Game Theory  Lab,  was composed of population of 47 persons (28 women and 19 men). The experimental game provided interesting data. A non-negligible proportion of the participants (35,71\% of women  and  36,84\% of the men population) have fully confessed. Considering the whole population, 36,17\% have fully confessed and
the 19,14\% have fully denied.  In terms of partial confession behaviors,  0.46\% of the women population and 0,54\% of the men population have partially confessed. Considering the whole population, 0,45\% have partially confessed in this reproduction of Prisoner Dilemma game.
Regarding the distribution of women and men population at the Interpersonal Reactivity Index, the experimental results reveal that the dominated strategies of the classical game theory are not dominated any more when users's psychology is involved, and a significant level of cooperation is observed among the users who are positively partially empathetic. 

The next future lines of our work would be based on a possible creation and implementation of a new model of empathy measurement that may take into account a multi-faceted presence of different variables  (general attitude to risk, a general estimation of the different heuristics present in the person, how is internalized individually the model Imagine Self/Imagine other that leads to reciprocity, individual attitude to fairness). Our aim is to make of this model a possible and valid model of measurement related to every day life situations where empathy is playing a key role not only in engineering field but, also, in social, economic and institutional area.

Everyday life seems to be really different and far from laboratory situations, where all the concepts are built around a crystalized idea of what it is or it should be. The evolutionary lines are then taken into account in our research as an important way to get data from a longitudinal perspective.
The dissatisfaction for a simple instrument of testing empathy leads us to rethink, first of all, our next step in empathy measurement. It would take into account two lines, essentially:

\begin{enumerate}
\item  a combined series of instrument of measurement, that will consider a multidimensional level of empathy.
\item a possibility to test and retest the variable in different moments and ways (construct validity, test and retest the person about the same cluster, e. g affective empathy).
\item A feedback from verbal and not verbal communication analysis software.
\end{enumerate}

It would be interesting to investigate (i) if there is any relationship between age and strategic decision making,  (ii) How the altruism and spitefulness evolve with increasing age. Furthermore, gender difference (if any) should be investigated in a bigger population and in games with distribution-dependent payoffs \cite{temftg1,temftg2,temftg3,temftg4}.


\section*{Compliance with Ethical Standards}

Conflict of interest: The authors declare that they have no conflict of
interest.

Ethical approval: All procedures performed in studies involving human participants were in accordance with the ethical standards of the institutional and/or national research committee and with the 1964 Helsinki declaration and its later amendments or comparable ethical standards.

Informed consent: Informed consent was obtained from all individual participants included in the study.


\section{Statistical Data }

{\color{black} Below we provide statistical data from the experimental game conducted in the Laboratory.  

Tables \ref{table:PTfemNNt} \ref{table:ECfemNNsss} \ref{table:FSfemNN} \ref{table:PDfemNN} report the data for women in the four IRI scales (PT,EC,FS,PD).

Tables \ref{table:PTmenNN}, \ref{table:ECmenNN}, \ref{table:FSmenNN},\ref{table:PDmenNN} focus on men statistical data for the four IRI scales (PT,EC,FS,PD).
}

\begin{table}[htb]

\begin{minipage}{0.4\textwidth}
\centering

\begin{tabular}{|c|c|c|c|c|c|}
\hline
{ \bf { \color{black}  Women}} &  \multicolumn{5}{|c|}{   \bf { \color{black} positively  }   confess  } \\ \hline

{\color{black}  PT} & A  &  B & C &  D & E  \\ \hline
Woman 3 &0 & 3&0  & 4 &0 \\ \hline

Woman 5  & 0& 3& 3& 1& 0 \\ \hline

Woman 6 &0&1&2& 3&1 \\ \hline

Woman 11 &0 &2& 4& 1&0  \\ \hline

Woman 15 & 0 &  5& 2 &0  & 0  \\ \hline

Woman 16 &1&2 &1 &3&0    \\ \hline

Woman 17 & 0 &  3 &  3 & 0 & 1   \\ \hline

Woman 19 &  1& 1 & 0& 1 & 4  \\ \hline

Woman 24 & 2 & 1 & 2 & 2 &0 \\ \hline

Woman 27 & 0&  2 & 3 & 1 & 0  \\ \hline

\multicolumn{6}{|c|}{  } \\ \hline

Who is PT & \multicolumn{5}{|c|}{  3,6, 16, 19,24, 27 } \\ \hline
Not PT & \multicolumn{5}{|c|}{  5,11,15,17 } \\ \hline
Unclear & \multicolumn{5}{|c|}{ } \\ \hline
\end{tabular}
\end{minipage}

 \vspace{0.5 cm}

 \begin{minipage}{0.4\textwidth}
 \centering
 \begin{tabular}{|c|c|c|c|c|c|}
\hline
{ \bf { \color{black}  Women}} &  \multicolumn{5}{|c|}{  \bf { \color{black} partially  }   confess } \\ \hline

{\color{black}  PT} & A  &  B & C &  D & E  \\ \hline

Woman 2 &1 &3 & 3 & 0 &0  \\ \hline

Woman 4 &1 & 1& 2& 2&1  \\ \hline

Woman 7 &  2& 0 & 0&0& 5  \\ \hline

Woman 8 &2& 0& 2 & 3&0  \\ \hline

Woman 9 &  0&2 & 3 & 2 & 0  \\ \hline

Woman 10  & 0& 3& 2 & 2 &0 \\ \hline

Woman 12  &  0&  1&  5&  1& 0  \\ \hline

Woman 14  &  0&  1&  4&  2& 0  \\ \hline

Woman 18  &  0&  2&  0&  1& 4  \\ \hline

Woman 20  &  1&  1&  1&  1& 3  \\ \hline

Woman 21  &  1&  1&  1&  4& 0  \\ \hline

Woman 22  &  0&  0&  7&  0& 0  \\ \hline

Woman 23  &  0&  2&  3&  1& 1  \\ \hline

Woman 25  &  1&  3&  1&  0& 2  \\ \hline

Woman 26  &  0&  1&  3&  1& 1  \\ \hline
\multicolumn{6}{|c|}{  } \\ \hline

Who is PT & \multicolumn{5}{|c|}{ 4,7,8,9,10,12,14,18,20,21,22,23,26} \\ \hline
Not PT & \multicolumn{5}{|c|}{  2, 25 } \\ \hline
Unclear & \multicolumn{5}{|c|}{ } \\ \hline
\end{tabular}
\end{minipage}

 \vspace{0.5 cm}
 \begin{minipage}{0.4\textwidth}
 \centering
 
 \begin{tabular}{|c|c|c|c|c|c|}
\hline
{ \bf { \color{black}  Woman}} &  \multicolumn{5}{|c|}{ \bf Deny   } \\ \hline

{\color{black}  PT} & A  &  B & C &  D & E  \\ \hline

Woman 1 &0 &1 &2 &4 & 0 \\ \hline

Woman 13 &0 &1 &2 & 2 &2 \\ \hline

Woman 28 &  3 & 1 & 1 & 2 & 0\\ \hline

\multicolumn{6}{|c|}{  } \\ \hline

Who is PT & \multicolumn{5}{|c|}{ 1,13} \\ \hline
Not PT & \multicolumn{5}{|c|}{  28 } \\ \hline
Unclear & \multicolumn{5}{|c|}{ } \\ \hline
\end{tabular}
\end{minipage}

\caption[]{  { \bf PT: Woman Result (positively, partially and deny tables)}}
\label{table:PTfemNNt} 
\end{table}



\begin{table}[htb]

\begin{minipage}{0.4\textwidth}
\centering
\begin{tabular}{|c|c|c|c|c|c|}
\hline
{ \bf { \color{black}  Woman}} &  \multicolumn{5}{|c|}{ \bf { \color{black} positively  }   confess  } \\ \hline

{\color{black}  EC} & A  &  B & C &  D & E  \\ \hline
Woman 3 &2 & 3&  1& 1 &0 \\ \hline

Woman 5  & 2& 1& 0& 4& 0 \\ \hline

Woman 6 &0&4&1& 1&1 \\ \hline

Woman 11 &0 &4& 1& 1&1  \\ \hline

Woman 15 & 3 &  0& 2 &2  & 0  \\ \hline

Woman 16 & 3& 1 &2 &1&0    \\ \hline

Woman 17 & 2 &  1 &  1 & 3 & 0   \\ \hline

Woman 19 &  3& 1 & 1& 1 & 1  \\ \hline

Woman 24 & 0 & 4 & 1& 1 &1 \\ \hline

Woman 27 & 2&  2 & 2 & 1 & 0  \\ \hline

\multicolumn{6}{|c|}{  } \\ \hline

Who is EC & \multicolumn{5}{|c|}{ 5,15,17 } \\ \hline
Not EC & \multicolumn{5}{|c|}{   3,6,11,16,19,24,27 } \\ \hline
Unclear & \multicolumn{5}{|c|}{ } \\ \hline
\end{tabular}
\end{minipage}

 \vspace{0.5 cm}

\begin{minipage}{0.4\textwidth}
\centering

\begin{tabular}{|c|c|c|c|c|c|}
\hline
{ \bf { \color{black}  Woman}} &  \multicolumn{5}{|c|}{  \bf { \color{black} partially  }   confess } \\ \hline

{\color{black}  EC} & A  &  B & C &  D & E  \\ \hline

Woman 2 &1 &2 & 0 & 3 &1  \\ \hline

Woman 4 &1 & 2& 4& 0&0  \\ \hline

Woman 7 &  1& 2 & 0&1& 3  \\ \hline

Woman 8 &3& 0& 1 & 3&0  \\ \hline

Woman 9 &  1&2 & 0 & 4 & 0  \\ \hline

Woman 10  & 3& 1& 1 & 2 &0 \\ \hline

Woman 12  &  0&  4&  2&  1& 0  \\ \hline

Woman 14  &  0&  0&  4&  3& 0  \\ \hline

Woman 18  &  1&  0&  2&  1& 3  \\ \hline

Woman 20  &  3&  0&  0&  1& 3  \\ \hline

Woman 21  &  2&  1&  0&  3& 1  \\ \hline

Woman 22  &  1&  2&  0&  4& 0  \\ \hline

Woman 23  &  0&  2&  2&  3& 0  \\ \hline

Woman 25  &  4&  0&  1&  1& 1  \\ \hline

Woman 26  &  2&  1&  2&  0& 2  \\ \hline
\multicolumn{6}{|c|}{  } \\ \hline

Who is EC & \multicolumn{5}{|c|}{ 2,9,18,14,7,20,21,8,22,23 } \\ \hline
Not EC & \multicolumn{5}{|c|}{   4,10,12,25,26 } \\ \hline
Unclear & \multicolumn{5}{|c|}{ } \\ \hline
\end{tabular}
\end{minipage}

 \vspace{0.5 cm}
 
\begin{minipage}{0.4\textwidth}
\centering

\begin{tabular}{|c|c|c|c|c|c|}
\hline
{ \bf { \color{black}  Woman}} &  \multicolumn{5}{|c|}{ { \bf Deny   }} \\ \hline

{\color{black}  EC} & A  &  B & C &  D & E  \\ \hline

Woman 1 &1 &1 &4 &1 & 0 \\ \hline

Woman 13 &0 &2 &2 & 3 &0 \\ \hline

Woman 28 &  2 & 1 & 2 & 0 & 2\\ \hline

\multicolumn{6}{|c|}{  } \\ \hline

Who is EC & \multicolumn{5}{|c|}{ 13} \\ \hline
Not EC & \multicolumn{5}{|c|}{  1,28 } \\ \hline
Unclear & \multicolumn{5}{|c|}{ } \\ \hline
\end{tabular}
\end{minipage}
\caption[]{ { \bf EC: Woman Result (positively, partially and deny tables)}}
\label{table:ECfemNNsss}
\end{table}

\begin{table}[htb]

\begin{minipage}{0.4\textwidth}
\centering
\begin{tabular}{|c|c|c|c|c|c|}
\hline
{ \bf { \color{black}  Woman}} &  \multicolumn{5}{|c|}{\bf { \color{black} positively  }   confess  } \\ \hline

{\color{black}  FS} & A  &  B & C &  D & E  \\ \hline
Woman 3 &4 & 1&  2& 0 &0 \\ \hline

Woman 5  & 3& 0& 3& 1& 0 \\ \hline

Woman 6 &0&2&4& 1&0 \\ \hline

Woman 11 &0 &2& 1& 4&0  \\ \hline

Woman 15 & 3 &  2& 2 &0  & 0  \\ \hline

Woman 16 & 1& 5 &0 &1&0    \\ \hline

Woman 17 & 2 &  1 &  2 & 2 & 0   \\ \hline

Woman 19 &  2& 3 & 0& 1 & 1  \\ \hline

Woman 24 & 0 & 1 & 5& 1 &0 \\ \hline

Woman 27 & 2&  0 & 3 & 2 & 0  \\ \hline

\multicolumn{6}{|c|}{  } \\ \hline

Who is FS & \multicolumn{5}{|c|}{ 11,5,6,17,24,3 } \\ \hline
Not FS & \multicolumn{5}{|c|}{  3,15,16,19 } \\ \hline
Unclear & \multicolumn{5}{|c|}{ } \\ \hline
\end{tabular}
\end{minipage}

 \vspace{0.5 cm}

\begin{minipage}{0.4\textwidth}
\centering
\begin{tabular}{|c|c|c|c|c|c|}
\hline
{ \bf { \color{black}  Woman}} &   \multicolumn{5}{|c|}{  \bf { \color{black} partially  }   confess } \\ \hline

{\color{black}  FS} & A  &  B & C &  D & E  \\ \hline

Woman 2 &2 &1 & 1 & 1 &2  \\ \hline

Woman 4 &1 & 4& 1& 0&1  \\ \hline

Woman 7 &  0& 3 & 1&3& 0  \\ \hline

Woman 8 &1& 1& 3 & 1&1  \\ \hline

Woman 9 &  2&0 & 0 & 5 & 0  \\ \hline

Woman 10  & 1& 6& 0 & 0 &0 \\ \hline

Woman 12  &  0&  1&  4&  2& 0  \\ \hline

Woman 14  &  0&  0&  2&  4& 1  \\ \hline

Woman 18  &  1&  1&  2&  2& 1  \\ \hline

Woman 20  &  2&  1&  1&  1& 2  \\ \hline

Woman 21  &  1&  1&  3&  2& 0  \\ \hline

Woman 22  &  1&  1&  2&  3& 0  \\ \hline

Woman 23  &  0&  0&  1&  1& 5  \\ \hline

Woman 25  &  1&  1&  2&  2& 1  \\ \hline

Woman 26  &  0&  4&  3&  0& 0  \\ \hline
\multicolumn{6}{|c|}{  } \\ \hline

Who is FS & \multicolumn{5}{|c|}{ 9,12,14,18,21,22,23,25 } \\ \hline
Not FS & \multicolumn{5}{|c|}{ 4,10,26  } \\ \hline
Unclear & \multicolumn{5}{|c|}{ } \\ \hline
\end{tabular}

\end{minipage}

 \vspace{0.5 cm}

\begin{minipage}{0.4\textwidth}
\centering

\begin{tabular}{|c|c|c|c|c|c|}
\hline
{ \bf { \color{black}  Woman}} & \multicolumn{5}{|c|}{ { \bf Deny   }} \\ \hline

{\color{black}  FS} & A  &  B & C &  D & E  \\ \hline

Woman 1 &2 &3 &1 &1 & 0 \\ \hline

Woman 13 &0 &0 &0 & 3 &4 \\ \hline

Woman 28 &  4 & 0 & 0 & 2 & 1\\ \hline

\multicolumn{6}{|c|}{  } \\ \hline

Who is FS & \multicolumn{5}{|c|}{ 1,28 } \\ \hline
Not FS & \multicolumn{5}{|c|}{  13 } \\ \hline
Unclear & \multicolumn{5}{|c|}{ } \\ \hline
\end{tabular}
\end{minipage}
\caption[]{  { \bf FS:  Women's Result (positively, partially and deny tables)}}
\label{table:FSfemNN}
\end{table}



\begin{table}[htb]
\begin{minipage}{0.4\textwidth}
\centering
\begin{tabular}{|c|c|c|c|c|c|}
\hline
{ \bf { \color{black}  Woman}} &   \multicolumn{5}{|c|}{ \bf { \color{black} positively  }   confess  } \\ \hline
{\color{black}  PD} & A  &  B & C &  D & E  \\ \hline
Woman 3  &1 & 2&  2& 2 &0 \\ \hline

Woman 5  & 3& 0& 0& 4& 0 \\ \hline

Woman 6 &2&3&0& 4&0 \\ \hline

Woman 11 &0 & 4& 3& 0&0  \\ \hline

Woman 15 & 0 &  4& 2 &1  & 0  \\ \hline

Woman 16 & 1& 2 &3 &1&0    \\ \hline

Woman 17 & 3 &  3 &  1 & 0 & 0   \\ \hline

Woman 19 &  4& 1 & 0& 1 & 1  \\ \hline

Woman 24 & 1 & 2 & 3& 1 &0 \\ \hline

Woman 27 & 2&  3 & 2 & 0 & 0  \\ \hline

\multicolumn{6}{|c|}{  } \\ \hline

Who is PD & \multicolumn{5}{|c|}{ 3,6,11,15,16,17,4,27,24} \\ \hline
Not PD & \multicolumn{5}{|c|}{  5 } \\ \hline
Unclear & \multicolumn{5}{|c|}{ } \\ \hline
\end{tabular}
\end{minipage}

 \vspace{0.5 cm}

\begin{minipage}{0.4\textwidth}
\centering
\begin{tabular}{|c|c|c|c|c|c|}
\hline
{ \bf { \color{black}  Woman}} &  \multicolumn{5}{|c|}{  \bf { \color{black} partially  }   confess } \\ \hline

{\color{black}  PD} & A  &  B & C &  D & E  \\ \hline

Woman 2 & 3 &0 & 1 & 1 &2  \\ \hline

Woman 4 &1 & 1& 4& 0&1  \\ \hline

Woman 7 &  1& 2 & 1&2& 1  \\ \hline

Woman 8 &3& 1& 2 & 1&0  \\ \hline

Woman 9 &  0&4 & 3 & 0 & 0  \\ \hline

Woman 10  & 4& 2& 1 & 0 &0 \\ \hline

Woman 12  &  0&  4&  2&  1& 0  \\ \hline

Woman 14  &  0&  1&  5&  1& 0  \\ \hline

Woman 18  &  2&  2&  1&  2& 0  \\ \hline

Woman 20  &  1&  2&  3&  0& 1  \\ \hline

Woman 21  &  1&  5&  0&  1& 0  \\ \hline

Woman 22  &  0&  3&  2&  1& 1  \\ \hline

Woman 23  &  0&  0&  5&  2& 0  \\ \hline

Woman 25  &  2&  4&  0&  0& 1  \\ \hline

Woman 26  &  2&  3&  1&  0& 1  \\ \hline
\multicolumn{6}{|c|}{  } \\ \hline

Who is PD & \multicolumn{5}{|c|}{ 23, 2, 14, 22  } \\ \hline
Not PD & \multicolumn{5}{|c|}{ 4, 8, 9, 10, 12, 18, 20, 21, 25, 26  } \\ \hline
Unclear & \multicolumn{5}{|c|}{ } \\ \hline
\end{tabular}
\end{minipage}

 \vspace{0.5 cm}

\begin{minipage}{0.4\textwidth}
\centering
\begin{tabular}{|c|c|c|c|c|c|}   
\hline
{ \bf { \color{black}  Woman}} &  \multicolumn{5}{|c|}{ { \bf Deny   }} \\ \hline

{\color{black}  PD} & A  &  B & C &  D & E  \\ \hline

Woman 1 &2 &1 &1 &1 & 1 \\ \hline

Woman 13 &0 &0 &3 & 4 &0 \\ \hline

Woman 28 &  3 & 1 & 0 & 1 & 2\\ \hline

\multicolumn{6}{|c|}{  } \\ \hline

Who is PD & \multicolumn{5}{|c|}{ 1,28 } \\ \hline
Not PD & \multicolumn{5}{|c|}{  13 } \\ \hline
Unclear & \multicolumn{5}{|c|}{ } \\ \hline
\end{tabular}
\end{minipage}
\caption[]{  { \bf PD: Women Result (positively, partially and deny tables)}}
\label{table:PDfemNN}

\end{table}



\begin{table}[htb]

\begin{minipage}{0.4\textwidth}
\centering
\begin{tabular}{|c|c|c|c|c|c|}
\hline
{ \bf { \color{black}  Men}} & \multicolumn{5}{|c|}{ \bf { \color{black} positively  }   confess  } \\ \hline

{\color{black}  PT} & A  &  B & C &  D & E  \\ \hline

Man 1  &0 & 3&  2& 1 &1 \\ \hline

Man 4   & 0& 2& 1& 1& 3 \\ \hline

Man 8  &0&2&4& 1&0 \\ \hline

Man 10 & 0&1 &2& 4& 1  \\ \hline

Man 11 & 3 &  0& 1 &1  & 2  \\ \hline

Man 12 &1&1 &2 &3&0    \\ \hline

Man 19 & 0 &  1 &  1 & 4 & 1   \\ \hline

\multicolumn{6}{|c|}{  } \\ \hline

Who is PT & \multicolumn{5}{|c|}{4,10,12,11,19 } \\ \hline
Not PT & \multicolumn{5}{|c|}{1,8   } \\ \hline
Unclear & \multicolumn{5}{|c|}{ } \\ \hline
\end{tabular}
\end{minipage}

 \vspace{0.5 cm}

\begin{minipage}{0.4\textwidth}
\centering
\begin{tabular}{|c|c|c|c|c|c|}
\hline
{ \bf { \color{black} Men}} &   \multicolumn{5}{|c|}{  \bf { \color{black} partially  }   confess } \\ \hline

{\color{black}  PT} & A  &  B & C &  D & E  \\ \hline

Man 3 & 0 &3 & 1 & 3 &0  \\ \hline

Man 5  &0 & 3& 3& 0&0  \\ \hline

Man 6 &  1& 0 & 1&4& 1  \\ \hline

Man 7 & 3& 3& 0 & 0&1  \\ \hline

Man 13 &  2&5 & 0 & 0 & 0  \\ \hline

Man 14  & 4 & 0& 1 & 0 &2 \\ \hline

\multicolumn{6}{|c|}{  } \\ \hline

Who is PT & \multicolumn{5}{|c|}{ 7,13,14} \\ \hline
Not PT & \multicolumn{5}{|c|}{ 5,6  } \\ \hline
Unclear & \multicolumn{5}{|c|}{ } \\ \hline
\end{tabular}

\end{minipage}

 \vspace{0.5 cm}

\begin{minipage}{0.4\textwidth}
\centering
\begin{tabular}{|c|c|c|c|c|c|}
\hline
{ \bf { \color{black}  Men}} &  \multicolumn{5}{|c|}{ { \bf Deny   }} \\ \hline

{\color{black}  PT} & A  &  B & C &  D & E  \\ \hline

Man 2  &0 &3 &2 &0 & 2 \\ \hline

Man 9  &0 &1 &3 &3 & 0 \\ \hline

Man 15  &0 &2 &1 &4 & 0 \\ \hline

Man 16  &0 &4 &3 &0 & 0 \\ \hline

Man 17  &0 &3 &3 &1 & 0 \\ \hline

Man 18  &0 &4 &3 &0 & 0 \\ \hline

\multicolumn{6}{|c|}{  } \\ \hline

Who is PT & \multicolumn{5}{|c|}{ 9,15,17 } \\ \hline
Not PT & \multicolumn{5}{|c|}{2,16,18} \\ \hline
Unclear & \multicolumn{5}{|c|}{ } \\ \hline
\end{tabular}
\end{minipage}
\caption[]{  { \bf PT: Men's Result (positively, partially and deny tables)}}
\label{table:PTmenNN}
\end{table}


\begin{table}[htb]

\begin{minipage}{0.4\textwidth}
\centering
\begin{tabular}{|c|c|c|c|c|c|}
\hline
{ \bf { \color{black}  Men}} &  \multicolumn{5}{|c|}{ \bf { \color{black} positively  }   confess  } \\ \hline

{\color{black}  EC} & A  &  B & C &  D & E  \\ \hline

Man 1  &3 & 3&  0& 1 &0 \\ \hline

Man 4   & 3& 0& 2& 1& 1 \\ \hline

Man 8  &0&4&3& 0&0 \\ \hline

Man 10 & 0&2 &4& 0& 1  \\ \hline

Man 11 & 1 &  0& 1 &4  & 1  \\ \hline

Man 12 &3&1 &1 &2&0    \\ \hline

Man 19 & 1&  0 &  1 & 4 & 1   \\ \hline

\multicolumn{6}{|c|}{  } \\ \hline

Who is EC & \multicolumn{5}{|c|}{ 10,11,19} \\ \hline
Not EC & \multicolumn{5}{|c|}{1,8,12,4  } \\ \hline
Unclear & \multicolumn{5}{|c|}{ } \\ \hline
\end{tabular}
\end{minipage}

 \vspace{0.5 cm}
 
\begin{minipage}{0.4\textwidth}
\centering
\begin{tabular}{|c|c|c|c|c|c|}
\hline
{ \bf { \color{black} Men}} &  \multicolumn{5}{|c|}{  \bf { \color{black} partially  }   confess } \\ \hline

{\color{black}  EC} & A  &  B & C &  D & E  \\ \hline

Man 3 & 0 &5 & 2 & 0 &0  \\ \hline

Man 5  &1 & 1& 3& 0&2  \\ \hline

Man 6 &  1& 1 & 0&4& 1  \\ \hline

Man 7 & 1& 2& 4 & 0&0  \\ \hline

Man 13 & 0& 5 &2 & 0 & 0  \\ \hline

Man 14  & 0 & 1& 4 & 2 &0 \\ \hline

\multicolumn{6}{|c|}{  } \\ \hline

Who is EC & \multicolumn{5}{|c|}{ 5,6,14} \\ \hline
Not EC & \multicolumn{5}{|c|}{ 3,7,13  } \\ \hline
Unclear & \multicolumn{5}{|c|}{ } \\ \hline
\end{tabular}
\end{minipage}

\vspace{0.3 cm}

\begin{minipage}{0.4\textwidth}
\centering
\begin{tabular}{|c|c|c|c|c|c|}
\hline
{ \bf { \color{black}  Men}} &  \multicolumn{5}{|c|}{ { \bf Deny   }} \\ \hline
Man
{\color{black}  EC} & A  &  B & C &  D & E  \\ \hline

Man 2  &2 &2 &0 &2 & 1 \\ \hline

Man 9  &0 &3 &1 &3 & 0 \\ \hline

Man 15  &1 &3 &3 &0 & 0 \\ \hline

Man 16  &3 &1 &3 &0 & 0 \\ \hline

Man 17  &0 &3 &4 &0 & 0 \\ \hline

Man 18  &3 &2 &1 &1 & 0 \\ \hline

\multicolumn{6}{|c|}{  } \\ \hline

Who is EC & \multicolumn{5}{|c|}{ 9 } \\ \hline
Not EC & \multicolumn{5}{|c|}{2,15,16,17,18} \\ \hline
Unclear & \multicolumn{5}{|c|}{ } \\ \hline
\end{tabular}
\end{minipage}
\caption[]{  { \bf EC: Men's Result group 3}}
\label{table:ECmenNN}
\end{table}



\begin{table}[htb]

\begin{minipage}{0.4\textwidth}
\centering
\begin{tabular}{|c|c|c|c|c|c|}
\hline
{ \bf { \color{black}  Men}} & \multicolumn{5}{|c|}{ \bf { \color{black} positively  }   confess  } \\ \hline

{\color{black}  FS} & A  &  B & C &  D & E  \\ \hline

Man 1  &1 & 3&  1& 0 & 2 \\ \hline

Man 4   & 1& 2& 1& 0& 1 \\ \hline

Man 8  &1&1&1& 1&3 \\ \hline

Man 10 & 0& 2 &3 & 2&  0  \\ \hline

Man 11 & 3 &  1& 0 &0  & 3  \\ \hline

Man 12 &0&2 &4 &1&0    \\ \hline

Man 19 & 0&  2 &  0 & 4 & 1   \\ \hline

\multicolumn{6}{|c|}{  } \\ \hline

Who is FS & \multicolumn{5}{|c|}{ 8,10,19} \\ \hline
Not FS & \multicolumn{5}{|c|}{1, 4,11,12 } \\ \hline
Unclear & \multicolumn{5}{|c|}{ } \\ \hline
\end{tabular}
\end{minipage}

 \vspace{0.5 cm}


\begin{minipage}{0.4\textwidth}
\centering
\begin{tabular}{|c|c|c|c|c|c|}
\hline
{ \bf { \color{black} Men}} &  \multicolumn{5}{|c|}{  \bf { \color{black} partially  }   confess } \\ \hline

{\color{black}  FS} & A  &Man  B & C &  D & E  \\ \hline

Man 3 & 2 &5 & 0 & 0 &0  \\ \hline

Man 5  &0 & 0& 4& 1&2  \\ \hline

Man 6 &  4& 3 & 0&0& 0  \\ \hline

Man 7 & 4& 3& 0 & 0&0  \\ \hline

Man 13 & 5& 1 &1 & 0 & 0  \\ \hline

Man 14  & 0 & 1& 6 & 0 &0 \\ \hline

\multicolumn{6}{|c|}{  } \\ \hline

Who is FS & \multicolumn{5}{|c|}{ 3,6,7,13} \\ \hline
Not FS & \multicolumn{5}{|c|}{ 5,14 } \\ \hline
Unclear & \multicolumn{5}{|c|}{ } \\ \hline
\end{tabular}
\end{minipage}

 \vspace{0.5 cm}

\begin{minipage}{0.4\textwidth}
\centering
\begin{tabular}{|c|c|c|c|c|c|}
\hline
{   \bf { \color{black}  Men} } &  \multicolumn{5}{|c|}{ { \bf Deny   }} \\ \hline

{\color{black}  FS} & A  &  B & C &  D & E  \\ \hline

Man 2  &3 &1 &0 &3 & 0 \\ \hline

Man 9  &0 &2 &4 &1 & 0 \\ \hline

Man 15  &0 &3 &2 &1 & 1 \\ \hline

Man 16  &1 &3 &2 &1 & 0 \\ \hline

Man 17  &1 &1 &4 &1 & 0 \\ \hline

Man 18  &0 &4 &3 &0 & 0 \\ \hline

\multicolumn{6}{|c|}{  } \\ \hline
Who is FS & \multicolumn{5}{|c|}{ 9, 17 } \\ \hline
Not FS & \multicolumn{5}{|c|}{2,15,16,18} \\ \hline
Unclear & \multicolumn{5}{|c|}{ } \\ \hline
\end{tabular}
\end{minipage}
  \caption[]{  { \bf FS: Men's Result (positively, partially and deny tables)}}
  \label{table:FSmenNN}
  
\end{table}




\begin{table}[htb]

\begin{minipage}{0.4\textwidth}
\centering
\begin{tabular}{|c|c|c|c|c|c|}
\hline
{ \bf { \color{black}  Men}} &  \multicolumn{5}{|c|}{ \bf { \color{black} positively  }   confess  } \\ \hline

{\color{black}  PD} & A  &  B & C &  D & E  \\ \hline

Man 1  &2 & 2&  0& 3 & 0 \\ \hline

Man 4   & 2& 0& 2& 0& 3 \\ \hline

Man 8  &0&3&2& 2&0 \\ \hline

Man 10 & 0& 3 &3 & 1&  0  \\ \hline

Man 11 & 1 &  1& 1 &0  & 4  \\ \hline

Man 12 &3&2 &2 &0&0    \\ \hline

Man 19 & 2&  1 &  2 & 1 & 1   \\ \hline

\multicolumn{6}{|c|}{  } \\ \hline

Who is PD & \multicolumn{5}{|c|}{ 4,10,11} \\ \hline
Not PD & \multicolumn{5}{|c|}{ 1,8,10,12,19} \\ \hline
Unclear & \multicolumn{5}{|c|}{ } \\ \hline
\end{tabular}
\end{minipage}

 \vspace{0.5 cm}

\begin{minipage}{0.4\textwidth}
\centering
\begin{tabular}{|c|c|c|c|c|c|}
\hline
{ \bf { \color{black} Men}} &   \multicolumn{5}{|c|}{  \bf { \color{black} partially  }   confess } \\ \hline

{\color{black}  PD} & A  &  B & C &  D & E  \\ \hline

Man 3 & 0 &5 & 2 & 0 &0  \\ \hline

Man 5  &0 & 4& 1& 2&0  \\ \hline

Man 6 &  1& 0 & 4&2& 0  \\ \hline

Man 7 & 4& 1& 0 & 0&2  \\ \hline

Woman 13 & 1& 3 &3 & 0 & 0  \\ \hline

Woman 14 & 0 & 4 & 1& 1  &1 \\ \hline

\multicolumn{6}{|c|}{  } \\ \hline

Who is PD & \multicolumn{5}{|c|}{ 6} \\ \hline
Not PD & \multicolumn{5}{|c|}{3,5,7,13,14} \\ \hline
Unclear & \multicolumn{5}{|c|}{ } \\ \hline
\end{tabular}


\end{minipage}

 \vspace{0.5 cm}

\begin{minipage}{0.4\textwidth}
\centering
\begin{tabular}{|c|c|c|c|c|c|}
\hline
{ \bf { \color{black}  Men}} &   \multicolumn{5}{|c|}{ { \bf Deny   }} \\ \hline
{\color{black}  PD} & A  &  B & C &  D & E  \\ \hline
Man 2  &4 &1 &0 &1 & 1 \\ \hline

Man 9  &1 &2 &3 &1 & 0 \\ \hline

Man 15  &1 &2 &0 &4 & 0 \\ \hline

Man 16  &0 &7 &0 &0 & 0 \\ \hline

Man 17  &2 &0 &1 &4 & 0 \\ \hline

Man 18  &0 &2 &4 &1 & 0 \\ \hline
\multicolumn{6}{|c|}{  } \\ \hline
Who is PD & \multicolumn{5}{|c|}{ 9,15,17,18 } \\ \hline
Not PD & \multicolumn{5}{|c|}{2,16} \\ \hline
Unclear & \multicolumn{5}{|c|}{ } \\ \hline
\end{tabular}
\end{minipage}
\caption[]{  { \bf PD: Men's Result (positively, partially and deny tables)}}
\label{table:PDmenNN}
\end{table}
\newpage

\section*{Biography}

{\bf Giulia Rossi}  received  her Master degree  with summa cum laude  in  Clinical Psychology in 2009 from the University of Padova. She worked as independent researcher in the analysis and prevention of psychopathological diseases and in the intercultural expression of mental diseases.
Her research interests include behavioral game theory, social norms and the epistemic foundations of mean-field-type game theory. 	She is currently a research associate in the Learning \& Game Theory Laboratory at New York University Abu Dhabi.

{\bf Alain Tcheukam}  received  his PhD in   2013 in Computer Science and Engineering at the IMT Institute for Advanced Studies Lucca. 
His research interests include crowd flows, smart cities and mean-field-type optimization.
He received the  Federbim Valsecchi award 2015 for his contribution in design, modelling and analysis of smarter cities, and a best paper award 2016 from the International Conference on Electrical Energy and Networks. He is currently a postdoctoral researcher with Learning \& Game Theory Laboratory at New York University Abu Dhabi.

{\bf Hamidou Tembine} (S'06-M'10-SM'13) received his M.S. degree in Applied Mathematics from Ecole Polytechnique and his Ph.D. degree in Computer Science  from University of Avignon. His current research interests include evolutionary games, mean field stochastic games and applications.
 In 2014, Tembine received the IEEE ComSoc Outstanding Young Researcher Award for his promising research activities for the benefit of the society. He was the recipient of 7 best paper awards in the applications of game theory. Tembine is a prolific researcher and holds several scientific publications including magazines, letters, journals and conferences. He is author of the book on "distributed strategic learning for engineers" (published by CRC Press, Taylor \& Francis 2012), and co-author of the book "Game Theory and Learning in Wireless Networks" (Elsevier Academic Press). Tembine has been co-organizer of several scientific meetings on game theory in networking, wireless communications and smart energy systems.   He is a senior member of IEEE.

\end{document}